\documentclass[preprint2]{aastex}

\newcommand{\myemail}{tim.rawle@sciops.esa.int}

\usepackage{amsmath}

\usepackage{array}
\newcolumntype{$}{>{\global\let\currentrowstyle\relax}}
\newcolumntype{^}{>{\currentrowstyle}}
\newcommand{\rowstyle}[1]{\gdef\currentrowstyle{#1}%
  #1\ignorespaces
}

\shorttitle{[CII] and CO(1--0) emission maps of a system at $z=5.24$}
\shortauthors{Rawle et al.}

\begin{document}

\title{[CII] and $^{12}$CO(1--0) Emission Maps in HLSJ091828.6+514223:\\ A Strongly Lensed Interacting System at $z=5.24$}


\author{T.~D.~Rawle\altaffilmark{1,2}, E.~Egami\altaffilmark{2}, R.~S.~Bussmann\altaffilmark{3}, M.~Gurwell\altaffilmark{3}, R.~J.~Ivison\altaffilmark{4}, F.~Boone\altaffilmark{5}, F.~Combes\altaffilmark{6}, A.~L.~R.~Danielson\altaffilmark{7}, M.~Rex\altaffilmark{2}, J.~Richard\altaffilmark{8}, I.~Smail\altaffilmark{7}, A.~M.~Swinbank\altaffilmark{7}, B.~Altieri\altaffilmark{1}, A.~W.~Blain\altaffilmark{9},  B.~Clement\altaffilmark{2}, M.~Dessauges-Zavadsky\altaffilmark{10}, A.~C.~Edge\altaffilmark{7}, G.~G.~Fazio\altaffilmark{3}, T.~Jones\altaffilmark{11}, J.-P.~Kneib\altaffilmark{12}, A.~Omont\altaffilmark{13}, P.~G.~P\'{e}rez-Gonz\'{a}lez\altaffilmark{14,2}, D.~Schaerer\altaffilmark{10,15}, I.~Valtchanov\altaffilmark{1}, P.~P.~van~der~Werf\altaffilmark{16}, G.~Walth\altaffilmark{2}, M.~Zamojski\altaffilmark{10}, M.~Zemcov\altaffilmark{17,18}}

\email{\myemail}

\altaffiltext{1}{ESAC, ESA, PO Box 78, Villanueva de la Ca\~{n}ada, 28691 Madrid, Spain}
\altaffiltext{2}{Steward Observatory, University of Arizona, 933 N. Cherry Ave, Tucson, AZ 85721}
\altaffiltext{3}{Harvard-Smithsonian Center for Astrophysics, 60 Garden Street, Cambridge, MA 02138}
\altaffiltext{4}{Institute for Astronomy, University of Edinburgh, Royal Observatory, Blackford Hill, Edinburgh EH9 3HJ, UK}
\altaffiltext{5}{Universit\'e de Toulouse, UPS-OMP, CNRS, IRAP, 9 Av. colonel Roche, BP 44346, 31028, Toulouse Cedex 4, France}
\altaffiltext{6}{Observatoire de Paris, LERMA, CNRS, 61 Av. de l'Observatoire, 75014, Paris, France}
\altaffiltext{7}{Institute for Computational Cosmology, Durham University, South Road, Durham, DH1 3LE, UK}
\altaffiltext{8}{CRAL, Universit\'e Lyon-1, 9 Av. Charles AndrŽ, 69561, St Genis Laval, France}
\altaffiltext{9}{Department of Physics \& Astronomy, University of Leicester, University Road, Leicester LE1 7RH, UK}
\altaffiltext{10}{Observatoire de Gen\`{e}ve, Universit\'{e} de Gen\`{e}ve, 51 Ch. des Maillettes, 1290, Sauverny, Switzerland}
\altaffiltext{11}{Department of Physics, University of California, Santa Barbara, CA 93106}
\altaffiltext{12}{Laboratoire d'Astrophysique EPFL, Observatoire de Sauverny, Versoix, 1290, Switzerland}
\altaffiltext{13}{13 Institut d'Astrophysique de Paris, CNRS and UPMC Univ. Paris 06,
UMR7095, 98bis Bd Arago, F-75014, Paris, France}
\altaffiltext{14}{Departamento de Astrof\'{\i}sica, Facultad de CC. F\'{\i}sicas,Universidad Complutense de Madrid, E-28040 Madrid, Spain}
\altaffiltext{15}{CNRS, IRAP, 14 Avenue E. Belin, 31400 Toulouse, France}
\altaffiltext{16}{Leiden Observatory, Leiden University, PO Box 9513, 2300 RA Leiden, The Netherlands}
\altaffiltext{17}{California Institute of Technology, Pasadena, CA 91125}
\altaffiltext{18}{Jet Propulsion Laboratory, Pasadena, CA 91109}


\begin{abstract}
We present Submillimeter Array (SMA) [CII] 158 \micron{} and Karl G. Jansky Very Large Array (VLA) $^{12}$CO(1--0) line emission maps for the bright, lensed, submillimeter source at $z=5.2430$ behind Abell 773: HLSJ091828.6+514223 (HLS0918). We combine these measurements with previously reported line profiles, including multiple $^{12}$CO rotational transitions, [CI], water and [NII], providing some of the best constraints on the properties of the interstellar medium (ISM) in a galaxy at $z>5$. HLS0918 has a total far-infrared (FIR) luminosity $L_{\rm FIR (8-1000\mu m)}$ $=$ (1.6 $\pm$ 0.1) $\times$ 10$^{14}$ L$_\sun$/$\mu$, where the total magnification $\mu_{\rm total}=8.9\pm1.9$, via a new lens model from the [CII] and continuum maps. Despite a HyLIRG luminosity, the FIR continuum shape resembles that of a local LIRG. We simultaneously fit all of the observed spectral line profiles, finding four components which correspond cleanly to discrete spatial structures identified in the maps. The two most redshifted spectral components occupy the nucleus of a massive galaxy, with a source plane separation $<1$ kpc. The reddest dominates the continuum map (de-magnified $L_{\rm FIR,component} = (1.1\pm0.2)\times10^{13}$ L$_\sun$), and excites strong water emission in both nuclear components via a powerful FIR radiation field from the intense star formation. A third star-forming component is most likely a region of a merging companion ($\Delta V$ $\sim$ 500 km s$^{-1}$) exhibiting generally similar gas properties. The bluest component originates from a spatially distinct region, and photo-dissociation region (PDR) analysis suggests that it is lower density, cooler and forming stars less vigorously than the other components. Strikingly, it has very strong [NII] emission which may suggest an ionized, molecular outflow. This comprehensive view of gas properties and morphology in HLS0918 previews the science possible for a large sample of high-redshift galaxies once ALMA attains full sensitivity.
\end{abstract}


\section{Introduction}
\label{sec:intro}

Atomic and molecular lines are particularly important diagnostics of star formation, active galactic nuclei (AGN) and the ISM, as the properties of the gas are dictated by the heating and cooling of various species. Furthermore, spectrally-resolved emission line observations provide a direct insight into the dynamics of a system.

The observed FIR and submillimeter (submm) spectrum of an actively star-forming galaxy is dominated by emission lines originating in photo-dissociation regions (PDRs), the surfaces of molecular clouds exposed to the ionizing UV flux from nearby young stars. Carbon is both abundant and possesses a lower ionization potential than hydrogen (11.3 eV compared to 13.6 eV), making the [CII] 158 \micron{} line a very important coolant for regions with neutral hydrogen. Although [CII] principally traces the UV-irradiated molecular clouds, the line is largely unaffected by extinction in plausible local environments (although see discussion in e.g. \citealt{ger98-17}), making it an excellent diagnostic for the physical conditions of the neutral gas \citep[e.g.][]{mai05-51,ven12-25}. Any interpretation must also acknowledge that significant [CII] emission (as much as 50\%) may arise from H{\sc ii} regions or cool, diffuse interstellar gas clouds \citep[e.g.][]{mad93-579}.

Whereas [CII] traces both neutral and ionized regions, the higher ionization potential of [NII] (14.53 eV) implies that its fine-structure lines trace only the ionized regime. In particular, the [NII] 205 \micron{} line has a very similar critical density to [CII] 158 \micron{}, meaning that [NII] is useful discriminant of the origin of the [CII] emission and the [NII]/[CII] ratio is relatively insensitive to gas properties such as density. In addition, the [NII]/[CII], ratio is observed to be proportional to $Z_{\rm gas}$ in Galactic HII regions \citep[e.g.][]{zee98-1}. At 158 and 205 \micron{}, [CII] and [NII] emission are less susceptible to absorption than other traditional metallicity indicators \citep{nag12-34}. However, nitrogen is a secondary element in gas enrichment, and at low metallicity or at high redshift, the gas may be enriched mostly by primary elements \citep{zee98-1,con02-75}, causing N/C to underestimate metallicity. The ability of the ratio to trace metallicity in high redshift star-forming regions, where the relative extent of [CII] and [NII] emission is unknown, is likely to be limited.

Molecular line emission from $^{12}$CO, observed as a series of rotational transitions, is another useful diagnostic for the properties of the ISM. In particular, $^{12}$CO(1--0) line intensity is used to estimate the molecular gas mass via a conversion factor $\alpha_{\rm CO}$ which relies on CO tracing the underlying H$_2$ gas mass \citep[e.g.][]{sol05-677}. The $^{12}$CO(1--0) line itself is relatively faint, often forcing an extrapolation from a higher-$J$ transition, using uncertain conversion factors. The full spectral line energy distribution (SLED) has also proven a useful diagnostic of the underlying properties of the molecular gas reservoir \citep[such as density and temperature distribution; e.g.][]{wei05-45,wei07-955,pap10-775, dan11-1687}. Modern PDR models (\citealt{kau99-795,mei07-793}) account for density, temperature and time-dependent chemistry to predict emission line ratios for many species, including $^{12}$CO, [CII] and [CI]. Comprehensive observation of many lines from a single source may be interpreted using these models to investigate the typical characteristics of the underlying PDR.

At $z\sim2$, submillimeter galaxies (SMGs) are a significant site of star formation, contributing up to 50\% of the cosmic star formation rate (SFR) density \citep{cha05-772,war11-1479}. However, our understanding of the evolution of galaxies throughout cosmic history is dependent on constraining star formation in the earlier era of stellar mass assembly, with the first major epoch occurring at $z$ $\ga$ 5 \citep[e.g.][]{sta10-1628}. 

The negative slope of the Rayleigh--Jeans tail to the dust component spectral energy distribution (SED) provides a `negative FIR K-correction' as we observe at higher redshifts. Although this tends to aid detection of SMGs at $z>4$, current instrumentation still only allows detailed observation of the brightest, high-redshift SMGs \citep[e.g.][]{sch08-5,rie09-1338,cop10-103,swi12-1066,wal12-93,rie13-329}. However, strong gravitational lensing (by an individual foreground galaxy and/or a massive cluster) increases the signal-to-noise, providing access to a large suite of emission lines for intrinsically fainter or more distant sources \citep{bak04-125,cop07-936,dan11-1687,cox11-63,wei13-88}. Lensing also increases the apparent angular size of an image, revealing high-redshift sources in incredible sub-kpc spatial resolution \citep[e.g.][]{swi09-1121,jon10-1247}. Observed source size often correlates with wavelength in lensed sources, so tailored apertures are required for spatially-integrated analysis, and care must be taken when comparing maps from different spectral windows. We assume that the effect of differential magnification is always negligible, preserving line ratios. This is a good approximation for the bolometric fraction of [CII] and $^{12}$CO(1--0), but may introduce some distortion to the CO ladder: e.g. $^{12}$CO(6--5)/$^{12}$CO(1--0) may have uncertainties of up to $\sim$30\% due to differential magnification \citep[e.g.][]{bla99-261,ser12-2429}.

In \citet[][hereafter C12]{com12-4}, we presented spatially-integrated observations of the bright, lensed SMG HLSJ091828.6+514223 (hereafter HLS0918; RA $=$ 09:18:28.6, Dec $=$ +51:42:23). The SMG was discovered in deep FIR imaging from the ``Herschel Lensing Survey" \citep{ega10-12}, located at a projected distance of 5.6$'$ from the X-ray center of the massive cluster Abell 773 ($z=0.22$). An intermediate-redshift galaxy (at $z=0.63$) dominates optical imaging at the source position, and also contributes $\sim$90\% of the lensing effect. C12 reported a source redshift of $z=5.24$\footnote{We adopt a systemic redshift $z=5.2430$, as derived from the center of the broad red peak in the high signal-to-noise [CII] line. Note that C12 used $z=5.2429$ from $^{12}$CO(6--5).} and an estimated total magnification of $\mu\sim11$. Two kinematic components were identified in the line profiles of multiple $^{12}$CO transitions, [NII], [CI] and a water line: a generally brighter, lower-frequency component with water emission, suggesting the presence of an intense FIR radiation field, and a higher-frequency component most prominent in [NII], tentatively interpreted as an ionized gas flow. Recently, \citet{lev12-9} utilized the unique set of observed lines to derive a constraint on hypothetical temporal variations in the fundamental physical constants of the Universe.

In this paper, we present new [CII] and $^{12}$CO(1--0) line emission maps of HLS0918. These are used to characterize the gas properties and morphology of the system in detail.  We adopt a new model for the combined cluster/galaxy lens (see Section \ref{sec:lensing}) which predicts a total `flux-weighted' magnification factor of $\mu\sim9$ for HLS0918 and allows us to account for relative amplification of the various components. A detailed description of the lensing constraints and parameters will be presented in Boone et al. (in prep.).

The paper is organized as follows. In Section \ref{sec:obs} we present the new maps of HLS0918, and summarize existing observations. In Section \ref{sec:results} we analyze the source continuum, derive a self-consistent fit to all of the line profiles, and examine the spatially-resolved emission in the maps. Section \ref{sec:discussion} discusses possible physical interpretation of the results in terms of source-plane morphology and gas properties. Our conclusions are presented in Section \ref{sec:conc}. Throughout the paper, we use $\Lambda$CDM cosmology with H$_0$=72 km s$^{-1}$ Mpc$^{-1}$, $\Omega_m$=0.27 and $\Omega_\Lambda$=1--$\Omega_m$.

\section{Observations}
\label{sec:obs}

This section describes all observations of HLS0918 employed in the paper: updated reduction of the {\it Herschel} discovery maps; new, spatially-resolved, line emission maps of [CII] from SMA and $^{12}$CO(1--0) from VLA; new IRAM 30m continuum and line observations; additional line profiles and continuum imaging published in C12. We defer description of the line profiles and spatial configuration of the maps until Section \ref{sec:results}.

\subsection{{\it Herschel} and {\it Spitzer}}

HLS0918 was discovered as an unusually bright `500 \micron{} peaker' source towards the edge of {\it Herschel}/SPIRE \citep{pil10-1,gri10-3} maps (250, 350 and 500 \micron{}) of Abell 773, observed as part of the ``{\it Herschel} Lensing Survey" \citep{ega10-12}. The SPIRE observations use large map mode, with a coverage of $\sim$17$'$ $\times$ 17$'$. The images were produced via the standard reduction pipeline in HIPE v9.0, the SPIRE Photometer Interactive Analysis (SPIA) package v1.7, and v8.1 calibration product. Note that this is a significant re-reduction since C12, including an improved treatment of the baseline removal (also known as `de-striping'). The SPIRE images are confusion limited (3$\,\sigma$ $=$ 17, 19, 20 mJy; \citealt{ngu10-5}) with beam sizes of 18, 25, 36$''$ (250, 350, 500 \micron{} respectively). 

{\it Herschel}/PACS and {\it Spitzer}/IRAC and MIPS imaging is only available for the core of Abell 773, and do not cover HLS0918 at $r_{\rm cluster}=5.6'$.

\subsection{Submillimeter Array (SMA)}

\begin{figure*}
\centering
\includegraphics[scale=0.28]{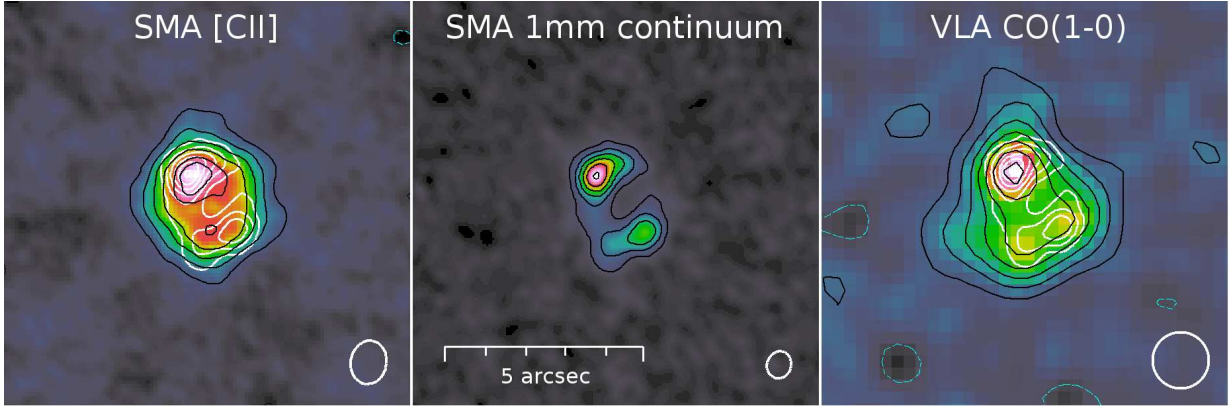}
\caption{SMA [CII] line emission ({\it left panel}), SMA 1mm (observed-frame) continuum ({\it center}) and VLA $^{12}$CO(1--0) line emission ({\it right}) maps showing $\sim10''\times10''$ region around HLS0918. North and east are aligned to the top and left, and the restoring beam size for each map is shown by a white ellipse in the lower-right corners. While the colormap is `min--max' scaled to accentuate structure, the thick black contours give the \{2,4,6...\}, \{5,10,15...\} and \{1,2,3...\} $\sigma$ confidence levels (left, center and right panels respectively. In each panel, thin dashed cyan lines show symmetric negative contours (\{--2,--4,--6...\}, \{--5,--10,--15...\} and \{--1,--2,--3...\} $\sigma$ respectively). For reference, the morphology of the continuum map is replicated in the other panels as white contours.}
\label{fig:maps}
\end{figure*}

SMA imaging data for the [CII] emission line of HLS0918 was obtained in the December 2011 (compact array), February 2012 (extended array) and April 2012 (very extended array), reaching a maximum baseline length of $\sim$500 m. Observations included $\sim$7 hours of on-source integration time per configuration and were conducted in superb weather conditions ($\tau_{\rm 225 GHz}\approx0.05$, phase errors between 10--30 degrees rms). We used the SMA single-polarization 345 GHz receivers, which provide an intermediate frequency coverage of 4Ð8 GHz, totaling 4 GHz bandwidth per sideband. The receivers were tuned such that the local oscillator frequency was 299.20478 GHz. This ensured that the upper sideband included the entire redshifted [CII] line at $\lambda_{\rm rest} = 157.74090$ \micron{}, while the lower sideband covered line-free continuum emission from the source at $\lambda_{\rm rest} \sim 160$ \micron{} ($\lambda_{\rm obs} \sim 1$ mm).

Calibration of the {\it uv} visibilities was performed using the Interactive Data Language (IDL) {\sc MIR} package. The blazar 3C84 was used as the primary bandpass calibrator and Titan was used for absolute flux calibration.  The nearby quasars 0927$+$390 ($F_{\rm 880 \mu m} = 0.5$ Jy, 6 degrees from target) and 0920$+$446 ($F_{\rm 880 \mu m} = 2.2$ Jy, 15 degrees from target) were used for phase and amplitude gain calibration, respectively.

The Multichannel Image Reconstruction, Image Analysis, and Display (MIRIAD) software package \citep{sau95-433} was used to invert the {\it uv} visibilities and deconvolve the dirty map.  Natural weighting was chosen to obtain maximum sensitivity and resulted in an elliptical Gaussian beam with a full-width half-maximum (FWHM) of 1.33$''$ $\times$ 1.10$''$ and position angle of 94 degrees east of north. The final integrated (along the frequency axis; 1$\,\sigma\sim1.1$~mJy/beam) [CII] emission map is shown in the left panel of Figure \ref{fig:maps}.

The final 1mm continuum map was produced from the compact and very extended configurations using {\sc aips} with an {\sc imarg robust} factor of 0.0 and a slight down-weighting of the shortest baselines to increase spatial resolution and emphasize structure. This scheme achieves a synthesized beam with 0.69$''$ $\times$ 0.60$''$ FWHM and a sensitivity limit of $\sim$0.6 mJy/beam. The map is displayed in the central panel of Figure \ref{fig:maps}.

This paper also uses SMA continuum observations of the source obtained at 880 \micron{} (341 GHz) and 1.3 mm (235 GHz) in the compact configuration. The beam sizes for these maps are 2.1$''$ $\times$ 2.0$''$ FWHM and 4.6$''$ $\times$ 2.8$''$ FWHM, respectively, which just resolves the source in the former case. These observations were first presented in C12.

\subsection{Jansky Very Large Array (VLA)}

We acquired data using the National Radio Astronomy Observatory (NRAO) VLA in the DnC and C configurations, during January and April 2012. For approximately 5 hours, we observed two near-contiguous sets of eight contiguous baseband pairs, each comprising $64\times 2$-MHz channels for a total dual-polarization bandwidth of 2,048 MHz. We tuned near the $^{12}$CO(1--0) transition \citep[$\nu_{\rm rest}=115.271203$\,GHz,][]{mor94-301} at $\nu_{\rm obs}=18.464$ GHz, having offset by 64 MHz to avoid the baseband edge.

Short slots, typically 2 hr long, were scheduled dynamically to ensure excellent phase stability and transparency in the K atmospheric window. The bright ($S_{\rm 18.5GHz}\approx 0.8$ Jy) calibration source, J0937+5008, was observed every few minutes to determine accurate complex gain solutions and bandpass corrections. 3C 286 was observed to set the absolute flux scale, and the pointing accuracy was checked locally every hour.

The data were reduced and imaged using {\sc aips} following the recipes described by \citet{ivi11-1913}, though with a number of significant changes: data were loaded using {\sc bdf2aips}, avoiding any compression, and {\sc fring} was used to optimize the delays, in software, based on 1 min of data for 3C 286. The basebands were knitted together using the {\sc noifs} task, yielding a $uv$ dataset with 1024 $\times$ 2-MHz channels centered near the expected frequency of the CO line. The channels were then imaged in groups of four (yielding 130 km s$^{-1}$ velocity resolution), with natural weighting ({\sc robust = 5}), to form a cube with spatial resolution, $\sim$1.4$''$, and a near-circular beam. The map, integrated along the frequency axis (1$\,\sigma\sim0.1$~mJy/beam), is shown in the right panel of Figure \ref{fig:maps}.

Approximately one hour of integration was previously obtained using VLA for the $^{12}$CO(2--1) transition at 36.9 GHz, as previously reported in C12.

\subsection{IRAM 30m and PdBI}

We present a new 2mm flux from the Goddard--IRAM Superconducting 2-Millimeter Observer (GISMO; \citealt{sta08-709}) on the IRAM 30m, observed in April 2013 with an average $\tau_{225 GHz}\sim0.3$. The {\sc lissajous\_tiny} script was used to obtain 30 minutes on source, resulting in a $5'\times3'$ map with a beam FWHM of 17.5$''$ and a source peak signal-to-noise (SNR) of $\sim$10 (reduced via {\sc crush}; \citealt{kov08-45}). This observation forms part of extensive GISMO follow-up of bright, high-z sources discovered by HLS (IRAM program 221-12; PI: A. Edge) to be fully described in a separate paper. Upper limits on the continuum flux were also previously obtained at 2 and 3 mm using the EMIR heterodyne receivers mounted on the IRAM 30m, as described in C12.

Our analysis includes a new detection of the $^{12}$CO(8--7) transition, observed by EMIR in mid-December 2012. We tuned EMIR to observe a central frequency of 147.656GHz, and used the wobbler for an efficient subtraction of the sky emission. The atmospheric conditions were good, with 4--6 mm of precipitable water vapor and a system temperature of 130--150 K.

We also take advantage of existing IRAM observations, all presented in C12. These include IRAM 30m spectra for three additional high-$J$ $^{12}$CO transitions ($J_{\rm upper}$ = 5, 6, 7), [CI]($^3$P$_2$--$^3$P$_1$) 370 \micron{}, H$_2$O$_p$(2,0,2--1,1,1) 304 \micron{} and [NII] 205 \micron{}, and PdBI data for CO(7--6) and [CI].

\section{Continuum and line analysis}
\label{sec:results}

\begin{table}
\caption{HLS0918 observed continuum fluxes}
\label{tab:flux}
\begin{tabular}{ccc}
\tableline
\multicolumn{1}{c}{$\lambda_{\rm obs}$} & & $S_{\nu}$ \\
mm & & mJy \\
\tableline
0.25 & {\it Herschel}/SPIRE & 96 $\pm$ 8 \\
0.35 & {\it Herschel}/SPIRE & 179 $\pm$ 13 \\
0.50 & {\it Herschel}/SPIRE & 212 $\pm$ 15 \\
0.88 & SMA & 125 $\pm$ 8 \\
1.0 & SMA & 103 $\pm$ 9 \\
1.3 & SMA & 55 $\pm$ 7 \\
2 & IRAM-30m/EMIR & $<$21 \\
2 & IRAM-30m/GISMO & 9 $\pm$ 1 \\
3 & IRAM-30m/EMIR & $<$2 \\
\end{tabular}
\end{table}

\begin{figure}
\centering
\includegraphics[angle=270,scale=0.62]{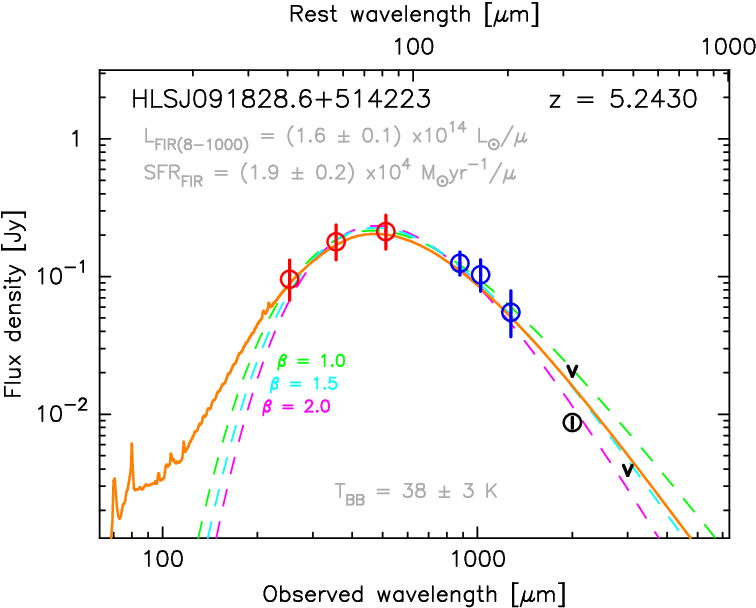}
\caption{{\it Herschel}/SPIRE (red), SMA (blue) and IRAM-30m (black) continuum fluxes. The dashed lines show the best fit modified blackbody (green, cyan and magenta for $\beta=1.0,1.5,2.0$ respectively). For $\beta=2.0$, which provides the best fit to the millimeter observations, $T_{\rm dust} = 38 \pm 3$ K. The solid (orange) line displays the best fit template from \citet{rie09-556}.}
\label{fig:fir}
\end{figure}

\subsection{FIR continuum and total luminosity}
\label{sec:lfir}

SPIRE photometry is derived via a two-dimensional Gaussian fit to the raw positions of the timestream data (HIPE routine {\sc timelineSourceFitterTask}), offering a robust measurement unaffected by the pixelization uncertainties inherent in map-making. These newly calculated SPIRE fluxes are $\sim10$ mJy larger than quoted in C12. These values, along with the new SMA 1mm continuum observation and all previous (sub)mm data, are listed in Table \ref{tab:flux}.

The updated infrared continuum is displayed in Figure \ref{fig:fir}, overlaid by the best fit \citet{rie09-556} template shape, allowing for an overall scaling to fit the source peak flux. The best fit template shape is derived from local LIRGs ($L_{\rm FIR; template}$ = 10$^{11.75}$ L$_\sun$). Integration under this template provides the far infrared luminosity of HLS0918, $L_{\rm FIR (8-1000\mu m)}$ = (1.6 $\pm$ 0.1) $\times$10$^{14}$ L$_\sun$ $\mu^{-1}$, where $\mu$ is the magnification factor (as displayed in the total system properties column of Table \ref{tab:derived}). Applying the \citet{ken98-189} relation, modified to match the initial mass function (IMF) of \citet{kro02-82}, as in \citet{rie09-556}\footnote{SFR [$M_{\sun}$ yr$^{-1}$] $=$ 0.66$\times$SFR$_{\rm Kennicutt}$ $=$ 1.138$\times$10$^{-10}$ $L_{\rm FIR}$ [L$_\sun$]}, this corresponds to a star-formation rate SFR = (1.9 $\pm$ 0.2) $\times$10$^{4}$ M$_\sun$ yr$^{-1}$ $\mu^{-1}$. If we instead assume the single-power law Salpeter IMF, as used by \citet{ken98-189}, the SFR would be $\sim$50\% higher.


We fit a single-temperature greybody to the FIR continuum in order to estimate the characteristic dust temperature ($T_{\rm dust}$) of the integrated source. We begin by fitting simultaneously for both $T_{\rm dust}$ and the dust emissivity index, $\beta$, resulting in $T_{\rm dust} = 48.7 \pm 3.3$ K and $\beta = 1.0 \pm 0.2$.\footnote{Following \citet{dac13-13}, we calculate a negligible dust heating contribution by the Cosmic Microwave Background (CMB; $T^{z=5.24}_{\rm CMB}$ $=$ 17 K) of $<$ 0.1 K.} However, the width of the fitted dust peak is driven by the SPIRE fluxes, with less weight given to the single GISMO flux and other IRAM limits, despite these millimeter points preferring a much narrower SED shape. Furthermore, although the above values have the maximum likelihood, a unique solution is not implied by the $\chi^2$ map in $T_{\rm dust}$--$\beta$ space, as one would expect from the well known degeneracy between these parameters \citep{bla99-261}. We therefore proceed with the common practice of assuming a typical value for $\beta$, which has the additional benefit of aiding comparisons with previous studies. We calculate the dust temperature assuming, in turn, both $\beta$ $=$ 1.5 and 2.0: $T_{\rm dust} = 42.6 \pm 2.9$ K and $T_{\rm dust} = 37.8 \pm 2.6$ K respectively, assuming that the adopted $\beta$ has an uncertainty of $\pm0.2$. The shape of the latter ($\beta=2.0$) is a better match for the millimeter data (see Figure \ref{fig:fir}), and agrees with derivations of $\beta$ for other ULIRGs \citep[e.g.][]{cle10-274}, so we adopt $38 \pm 3$ K as the canonical characteristic dust temperature of HLS0918.

The dust temperature is lower than those derived for the cores of local ULIRGs from single-temperature fits (e.g.  $T_{\rm dust}\approx$ 60--70 K for M82, Arp220; \citealt{pan10-37,ran11-94}), but consistent with typical IRAS galaxies ($\sim$40 K) and other SMGs observed by {\it Herschel} \citep[40--50 K, e.g.][]{dan11-1687,rie13-329}. Indeed, comparing to the general $L_{\rm FIR}$--$T_{\rm dust}$ relation \citep[][Figure 4]{hwa10-75}, and assuming a magnification factor of $\mu\sim10$, HLS0918 lies within the overlap between the typical ULIRG and SMG regions as expected. Even accounting for lensing, HLS0918 is an exceptionally bright source ($>$10$^{13}$ L$_\sun$, a `HyLIRG'), yet the shape of the dust continuum resembles a local LIRG, as reported previously for other luminous, high-redshift sources \citep[e.g.][]{rex10-13}.

\begin{table*}
\caption{Derived properties for HLS0918. Values are uncorrected for magnification $\mu$, unless explicitly stated. Results are presented for both the integrated source (bold `Total' column) and constituent components (Ra, Rb, B, VB; see Section \ref{sec:lines}). Component $L_{\rm FIR}$ assumes relative luminosities derived from the 1mm continuum map (presented in the first row), given the association between spectral and spatial components identified in Section \ref{sec:maps}.}
\label{tab:derived}
\begin{tabular}{lr>{\bfseries}cccccc}
\tableline
Parameter & \multicolumn{1}{c}{unit} & Total & \multicolumn{1}{c}{Ra}  & \multicolumn{1}{c}{Rb} & \multicolumn{1}{c}{B} & \multicolumn{1}{c}{VB} & \\
\tableline
Relative $L_{\rm 1mm}$ & & 1.00 $\pm$ 0.04 & 0.64 $\pm$ 0.03 & 0.14 $\pm$ 0.02 & 0.21 $\pm$ 0.02 & 0.02 $\pm$ 0.01 & (1) \\
\tableline
$T_{\rm dust}$ & K & 38 $\pm$ 3 & & & & & (2) \\
$L_{\rm FIR (8-1000\mu m)}$ & $\times$10$^{12}$ L$_\sun$ $\mu^{-1}$ & 160 $\pm$ 10 & 104 $\pm$ 8 & 23 $\pm$ 4 & 34 $\pm$ 4 & 2 $\pm$ 1 & (3) \\
SFR & $\times$10$^{3}$ M$_\sun$ yr$^{-1}$ $\mu^{-1}$ & 19 $\pm$ 2 & 12 $\pm$ 1 & 2.6 $\pm$ 0.5 & 4.0 $\pm$ 0.8 & 0.3 $\pm$ 0.2 & (3) \\
$L_{\rm FIR (42.5-122.5\mu m)}$ & $\times$10$^{12}$ L$_\sun$ $\mu^{-1}$ & 100 $\pm$ 6 & 65 $\pm$ 5 & 14 $\pm$ 2 & 21 $\pm$ 2 & 2 $\pm$ 1 & (3) \\
\tableline
$\mu$ & & 8.9 $\pm$ 1.9 & 9.4 $\pm$ 2.0 & 12.0 $\pm$ 2.6 & 8.4 $\pm$ 1.7 & 4.4 $\pm$ 0.9 & (4) \\
$L_{\rm FIR (8-1000\mu m), demag}$ & $\times$10$^{12}$ L$_\sun$ & 18 $\pm$ 4 & 11 $\pm$ 2 & 1.9 $\pm$ 0.4 & 4.0 $\pm$ 0.8 & 0.5 $\pm$ 0.3 & (4) \\
\tableline
$L_{\rm [CII]}/L_{\rm FIR}$ & $\times$10$^{-4}$ & 8.4 $\pm$ 0.5 & 5 $\pm$ 1 & 15 $\pm$ 5 & 10 $\pm$ 2 & 73 $\pm$ 38 & (5) \\
$L_{\rm CO(1-0)}/L_{\rm FIR}$ & $\times$10$^{-7}$ & 4.3 $\pm$ 0.4 & 3.1 $\pm$ 0.8 & 8.3 $\pm$ 3.7 & 4.7 $\pm$ 1.1 & 20 $\pm$ 20 & (5) \\
$M_{\rm gas}$ & $\times$10$^{9}$ M$_\sun$ & $\ga$74 $\pm$ 17 & $\ga$31 $\pm$ 10 & $\ga$15 $\pm$ 7 & $\ga$18 $\pm$ 5 & $\ga$10 $\pm$ 9 & (6) \\
SFE & L$_\sun$ M$_\sun^{-1}$ & $\la$250 $\pm$ 20 & $\la$360 $\pm$ 90 & $\la$130 $\pm$ 50 & $\la$230 $\pm$ 50 & $\la$60 $\pm$ 60 & (6) \\
$L_{\rm [NII]}/L_{\rm [CII]}$ & $\times$10$^{-2}$ & 5.2 $\pm$ 0.6 & 3.9 $\pm$ 1.6 & 4.7 $\pm$ 2.3 & 3.9 $\pm$ 1.7 & 12 $\pm$ 7 & \\
\tableline
\tableline
\end{tabular}

(1) Relative apparent 1mm (observed frame) continuum luminosity, Section \ref{sec:maps}\\
(2) From a modified blackbody fit, Section \ref{sec:lfir}\\
(3) From the best fit \citet{rie09-556} template, Section \ref{sec:lfir}\\
(4) From lensing model, Section \ref{sec:lensing}\\
(5) For comparison with literature \citep[e.g.][]{sta10-957,bre11-8}, we use $L_{\rm FIR (42.5-122.5\mu m)}$\\
(6) See Section \ref{sec:gasmass}
\end{table*}

\subsection{Atomic and molecular line emission}
\label{sec:lines}

\begin{figure*}
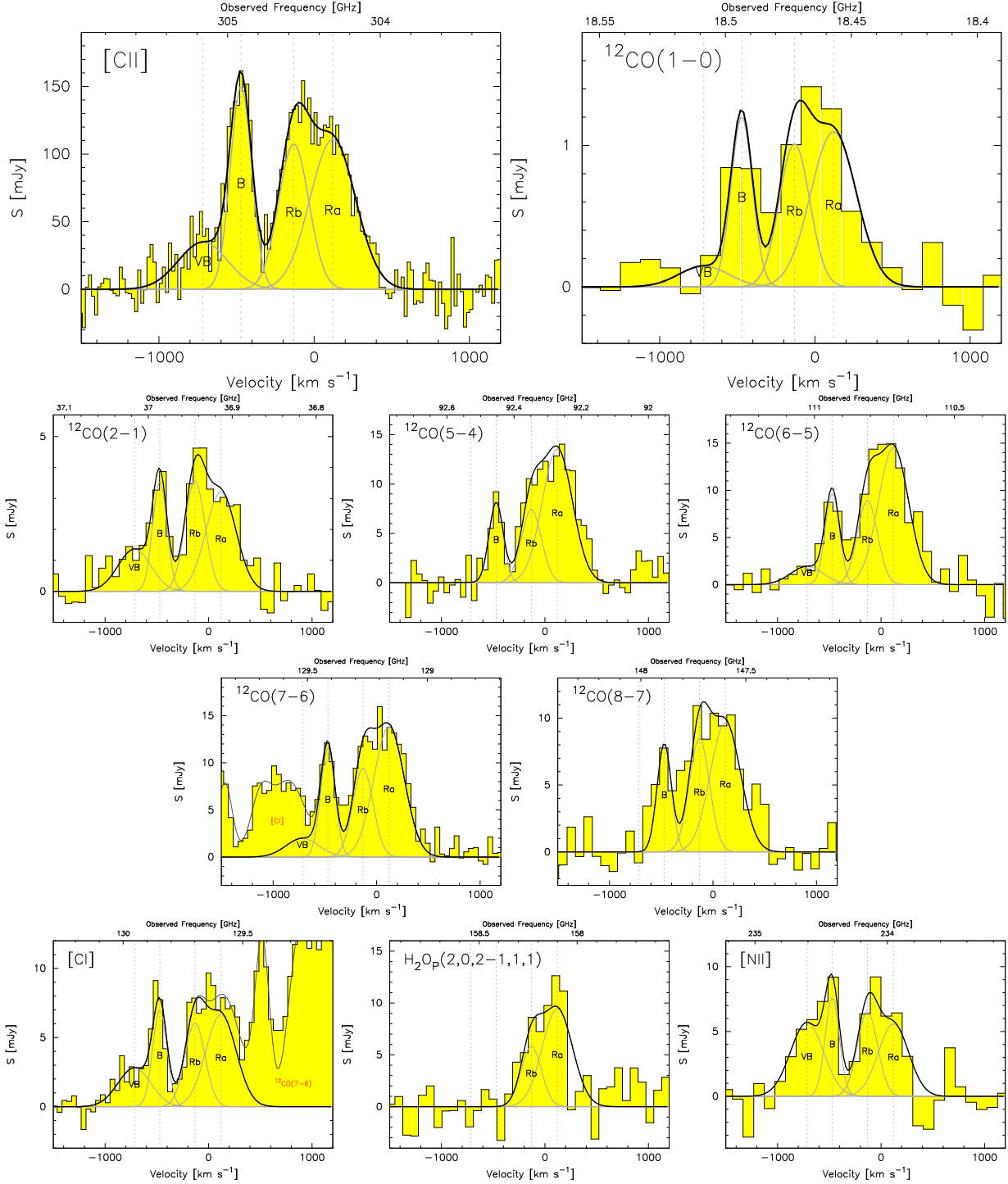

\centering
\includegraphics[angle=270,scale=0.51]{outspec_CII.eps}
\includegraphics[angle=270,scale=0.51]{outspec_CO10.eps} \\
\includegraphics[angle=270,scale=0.34]{outspec_CO21.eps}
\includegraphics[angle=270,scale=0.34]{outspec_CO54.eps}
\includegraphics[angle=270,scale=0.34]{outspec_CO65.eps} \\
\includegraphics[angle=270,scale=0.34]{outspec_CO76.eps}
\includegraphics[angle=270,scale=0.34]{outspec_CO87.eps} \\
\includegraphics[angle=270,scale=0.34]{outspec_CI.eps}
\includegraphics[angle=270,scale=0.34]{outspec_H2O.eps}
\includegraphics[angle=270,scale=0.34]{outspec_NII.eps}
\caption{Continuum-subtracted atomic and molecular line profiles: spatially-integrated profiles derived from the new [CII] and $^{12}$CO(1--0) maps are shown in the top row. The best, fully-simultaneous fit by four Gaussian-profile components are overlaid (Section \ref{sec:lines} for more details): two red (Ra and Rb; +120 $\pm$ 30 and --130 $\pm$ 20 km s$^{-1}$ respectively), a blue (B; --470 $\pm$ 10 km s$^{-1}$) and a `very blue' (VB; --720 $\pm$ 40 km s$^{-1}$). All velocities are relative to the center of the broad [CII] peak ($z=5.2430$). VB exhibits unusually strong [NII] emission, but no detection for $^{12}$CO(5--4) or $^{12}$CO(8--7). H$_2$O is only detected for Ra and Rb. Other than this water line, the Rb/B ratio is consistently $\sim$1.2$\pm$0.1.}
\label{fig:lines}
\end{figure*}

In Figure \ref{fig:lines} we show the atomic and molecular line emission from HLS0918, which is spatially-integrated in the case of the [CII] and $^{12}$CO(1--0) maps. The profiles clearly show intensity and kinematic structure, with a similar spectral shape in each line. Here, we describe how we decompose the profiles into a number of spectral components, which can be used to create channel maps in [CII] and $^{12}$CO, thus characterizing the source-plane configuration of HLS0918.

First we examine the high signal-to-noise [CII] profile from SMA to generate an initial model. Two components are immediately obvious: a broad peak ($V\sim0$ km s$^{-1}$) and a narrower, blue-shifted peak (B; $V\sim-500$ km s$^{-1}$). We find that the $\chi^2$ of the fit is improved with the addition of a fainter third component at an even greater velocity offset (VB; $V\sim-750$ km s$^{-1}$). Moreover, the broad component appears double-peaked, and treating it as two separate entities (Ra, Rb; $V\sim$ +120, --130 km s$^{-1}$) reduces the $\chi^2$ further. Our best fit model of the line comprises these four Gaussian-profile components.

In order to produce a self-consistent, physically-meaningful set of line luminosities, we perform a simultaneous fit to all the observed line profiles described in Section \ref{sec:obs} -- [CII], $^{12}$CO (J$_{\rm upper}$=1,2,5,6,7,8), [CI], H$_2$O$_p$(2,0,2-1,1,1) and [NII]. Continuum flux is subtracted from each spectrum using a simple linear interpolation from frequencies beyond the emission line ($V$ $<$ --2000 km s$^{-1}$; $V$ $>$ 1000 km s$^{-1}$). The central velocity and width of each component is only allowed to vary globally (i.e the same for each line profile), while the peak flux has full freedom for each individual component of every line. This model assumes that for every component, emission in each line originates from the same physical region, which should prove to be a reasonable approximation. The total number of free parameters in the fit is $N = N_{\rm C}(2+N_{\rm L})$ where $N_{\rm C}$ is the number of components and $N_{\rm L}$ is the number of lines. For our current data, $N$ $=$ 4$\times$(2+10) $=$ 48. Uncertainties are calculated via 2000 Monte Carlo simulations in which the observed spectra are randomly fluctuated by their 1$\,\sigma$ variance.

\begin{table*}
\caption{Derived properties for the observed atomic and molecular lines. All values are uncorrected for magnification.}
\label{tab:fits}
\begin{tabular}{$l^l^c^c^c^c^c^c}
\tableline
 & & \multicolumn{1}{c}{$\nu_{\rm obs}$} & \multicolumn{1}{c}{$S_\nu$} & \multicolumn{1}{c}{$\Delta V_{\rm FWHM}$} & \multicolumn{1}{c}{$V$} & \multicolumn{1}{c}{$I$} & \multicolumn{1}{c}{$L$} \\
Line & Component & \multicolumn{1}{c}{GHz} & \multicolumn{1}{c}{mJy} & \multicolumn{1}{c}{km s$^{-1}$} & \multicolumn{1}{c}{km s$^{-1}$} & \multicolumn{1}{c}{Jy km s$^{-1}$} & \multicolumn{1}{c}{10$^{9}$ L$_\sun$} \\
\tableline
[CII] & Ra & 304.3051 & 111 $\pm$ 9 & 350 $\pm$ 50 & +120 $\pm$ 30 & 41 $\pm$ 8 & 33 $\pm$ 6 \\
{[}CII] & Rb & 304.5606 & 108 $\pm$ 22 & 230 $\pm$ 30 & --130 $\pm$ 20 & 26 $\pm$ 7 & 21 $\pm$ 6 \\
{[}CII] & B & 304.9062 & 151 $\pm$ 12 & 160 $\pm$ 10 & --470 $\pm$ 10 & 26 $\pm$ 4 & 21 $\pm$ 3 \\
{[}CII] & VB & 305.1557 & 35 $\pm$ 8 & 370 $\pm$ 170 & --720 $\pm$ 40 & 14 $\pm$ 5 & 11 $\pm$ 4 \\
\rowstyle{\bfseries}
[CII] & Total & & & & & 107 $\pm$ 3 & 85 $\pm$ 2 \\\noalign{\smallskip}
\hline
$^{12}$CO(1--0) & Ra & 18.4567 & 1.1 $\pm$ 0.2 & 350 $\pm$ 50 & +120 $\pm$ 30 & 0.40 $\pm$ 0.10 & 0.020 $\pm$ 0.005 \\
$^{12}$CO(1--0) & Rb & 18.4722 & 1.0 $\pm$ 0.3 & 230 $\pm$ 30 & --130 $\pm$ 20 & 0.25 $\pm$ 0.09 & 0.012 $\pm$ 0.005 \\
$^{12}$CO(1--0) & B & 18.4931 & 1.2 $\pm$ 0.2 & 160 $\pm$ 10 & --470 $\pm$ 10 & 0.21 $\pm$ 0.04 & 0.010 $\pm$ 0.002 \\
$^{12}$CO(1--0)& VB & 18.5083 & 0.1 $\pm$ 0.1 & 370 $\pm$ 170 & --720 $\pm$ 40 & 0.06 $\pm$ 0.06 & 0.003 $\pm$ 0.003 \\
\rowstyle{\bfseries}
$^{12}$CO(1--0) & Total & & & & & 0.92 $\pm$ 0.07 & 0.044 $\pm$ 0.003 \\\noalign{\smallskip}
\hline
$^{12}$CO(2--1) & Ra & 36.9127 & 3.2 $\pm$ 0.4 & 350 $\pm$ 50 & +120 $\pm$ 30 & 1.2 $\pm$ 0.3 & 0.11 $\pm$ 0.03 \\
$^{12}$CO(2--1) & Rb & 36.9437 & 3.6 $\pm$ 0.7 & 230 $\pm$ 30 & --130 $\pm$ 20 & 0.9 $\pm$ 0.2 & 0.09 $\pm$ 0.02 \\
$^{12}$CO(2--1) & B & 36.9856 & 3.5 $\pm$ 0.6 & 160 $\pm$ 10 & --470 $\pm$ 10 & 0.6 $\pm$ 0.1 & 0.06 $\pm$ 0.01 \\
$^{12}$CO(2--1) & VB & 37.0158 & 1.4 $\pm$ 0.3 & 370 $\pm$ 170 & --720 $\pm$ 40 & 0.5 $\pm$ 0.2 & 0.05 $\pm$ 0.02 \\
\rowstyle{\bfseries}
$^{12}$CO(2--1) & Total & & & & & 3.2 $\pm$ 0.2 & 0.31 $\pm$ 0.02 \\\noalign{\smallskip}
$^{12}$CO(5--4) & Ra & 92.2693 & 13.5 $\pm$ 1.0 & 350 $\pm$ 50 & +120 $\pm$ 30 & 5.0 $\pm$ 0.8 & 1.2 $\pm$ 0.2 \\
$^{12}$CO(5--4) & Rb & 92.3468 & 7.4 $\pm$ 2.7 & 230 $\pm$ 30 & --130 $\pm$ 20 & 1.8 $\pm$ 0.8 & 0.4 $\pm$ 0.2 \\
$^{12}$CO(5--4) & B & 92.4516 & 8.1 $\pm$ 1.5 & 160 $\pm$ 10 & --470 $\pm$ 10 & 1.4 $\pm$ 0.3 & 0.34 $\pm$ 0.06 \\
$^{12}$CO(5--4) & VB & 92.5272 & & 370 $\pm$ 170 & --720 $\pm$ 40 & $<$0.9 & $<$0.2 \\
\rowstyle{\bfseries}
$^{12}$CO(5--4) & Total & & & & & 8.2 $\pm$ 0.5 & 2.0 $\pm$ 0.1 \\\noalign{\smallskip}
$^{12}$CO(6--5) & Ra & 110.7154 & 14.5 $\pm$ 1.5 & 350 $\pm$ 50 & +120 $\pm$ 30 & 5.3 $\pm$ 1.1 & 1.6 $\pm$ 0.3 \\
$^{12}$CO(6--5) & Rb & 110.8084 & 8.9 $\pm$ 3.4 & 230 $\pm$ 30 & --130 $\pm$ 20 & 2.2 $\pm$ 1.0 & 0.6 $\pm$ 0.3 \\
$^{12}$CO(6--5) & B & 110.9341 & 9.7 $\pm$ 1.9 & 160 $\pm$ 10 & --470 $\pm$ 10 & 1.7 $\pm$ 0.3 & 0.5 $\pm$ 0.1 \\
$^{12}$CO(6--5) & VB & 111.0249 & 1.9 $\pm$ 1.3 & 370 $\pm$ 170 & --720 $\pm$ 40 & 0.8 $\pm$ 0.5 & 0.2 $\pm$ 0.2 \\
\rowstyle{\bfseries}
$^{12}$CO(6--5) & Total & & & & & 10.0 $\pm$ 0.7 & 2.9 $\pm$ 0.2 \\\noalign{\smallskip}
$^{12}$CO(7--6) & Ra & 129.1573 & 13.8 $\pm$ 0.9 & 350 $\pm$ 50 & +120 $\pm$ 30 & 5.1 $\pm$ 0.8 & 1.7 $\pm$ 0.3 \\
$^{12}$CO(7--6) & Rb & 129.2657 & 9.4 $\pm$ 2.6 & 230 $\pm$ 30 & --130 $\pm$ 20 & 2.3 $\pm$ 0.8 & 0.8 $\pm$ 0.3 \\
$^{12}$CO(7--6) & B & 129.4124 & 11.6 $\pm$ 1.1 & 160 $\pm$ 10 & --470 $\pm$ 10 & 2.0 $\pm$ 0.3 & 0.7 $\pm$ 0.1 \\
$^{12}$CO(7--6) & VB & 129.5182 & 2.0 $\pm$ 1.1 & 370 $\pm$ 170 & --720 $\pm$ 40 & 0.8 $\pm$ 0.4 & 0.3 $\pm$ 0.2 \\
\rowstyle{\bfseries}
$^{12}$CO(7--6) & Total & & & & & 10.2 $\pm$ 0.4 & 3.5 $\pm$ 0.1 \\\noalign{\smallskip}
$^{12}$CO(8--7) & Ra & 147.5943 & 9.7 $\pm$ 1.1 & 350 $\pm$ 50 & +120 $\pm$ 30 & 3.6 $\pm$ 0.7 & 1.4 $\pm$ 0.3 \\
$^{12}$CO(8--7) & Rb & 147.7182 & 8.5 $\pm$ 2.1 & 230 $\pm$ 30 & --130 $\pm$ 20 & 2.1 $\pm$ 0.7 & 0.8 $\pm$ 0.3 \\
$^{12}$CO(8--7) & B & 147.8858 & 8.1 $\pm$ 1.5 & 160 $\pm$ 10 & --470 $\pm$ 10 & 1.4 $\pm$ 0.3 & 0.5 $\pm$ 0.1 \\
$^{12}$CO(8--7) & VB & 148.0068 & & 370 $\pm$ 170 & --720 $\pm$ 40 & $<$1.2 & $<$0.4 \\
\rowstyle{\bfseries}
$^{12}$CO(8--7) & Total & & & & & 7.4 $\pm$ 0.5 & 4.6 $\pm$ 0.3 \\\noalign{\smallskip}
{[}CI] & Ra & 129.5883 & 6.7 $\pm$ 0.9 & 350 $\pm$ 50 & +120 $\pm$ 30 & 2.5 $\pm$ 0.5 & 0.8 $\pm$ 0.2 \\
{[}CI] & Rb & 129.6971 & 6.1 $\pm$ 1.6 & 230 $\pm$ 30 & --130 $\pm$ 20 & 1.5 $\pm$ 0.5 & 0.4 $\pm$ 0.1 \\
{[}CI] & B & 129.8443 & 7.0 $\pm$ 1.3 & 160 $\pm$ 10 & --470 $\pm$ 10 & 1.2 $\pm$ 0.3 & 0.4 $\pm$ 0.1 \\
{[}CI] & VB & 129.9504 & 2.8 $\pm$ 0.8 & 370 $\pm$ 170 & --720 $\pm$ 40 & 1.1 $\pm$ 0.5 & 0.4 $\pm$ 0.2 \\
\rowstyle{\bfseries}
[CI] & Total & & & & & 6.3 $\pm$ 0.3 & 2.1 $\pm$ 0.1 \\\noalign{\smallskip}
H$_2$O & Ra & 158.0979 & 10 $\pm$ 2 & 350 $\pm$ 50 & +120 $\pm$ 30 & 3.5 $\pm$ 0.9 & 1.4 $\pm$ 0.3  \\
H$_2$O & Rb & 158.2306 & 6 $\pm$ 3 & 230 $\pm$ 30 & --130 $\pm$ 20 & 1.4 $\pm$ 0.8 & 0.6 $\pm$ 0.3  \\
\rowstyle{\bfseries}
H$_2$O & Total & & & & & 4.9 $\pm$ 0.6 & 2.0 $\pm$ 0.3  \\\noalign{\smallskip}
{[}NII] & Ra & 233.9499 & 6 $\pm$ 2 & 350 $\pm$ 50 & +120 $\pm$ 30 & 2.1 $\pm$ 0.7 & 1.3 $\pm$ 0.4 \\
{[}NII] & Rb & 234.1464 & 7 $\pm$ 2 & 230 $\pm$ 30 & --130 $\pm$ 20 & 1.6 $\pm$ 0.7 & 1.0 $\pm$ 0.4 \\
{[}NII] & B & 234.4121 & 8 $\pm$ 3 & 160 $\pm$ 10 & --470 $\pm$ 10 & 1.3 $\pm$ 0.5 & 0.8 $\pm$ 0.3 \\
{[}NII] & VB & 234.6039 & 6 $\pm$ 2 & 370 $\pm$ 170 & --720 $\pm$ 40 & 2.3 $\pm$ 0.9 & 1.4 $\pm$ 0.6 \\
\rowstyle{\bfseries}
[NII] & Total & & & & & 7.2 $\pm$ 0.8 & 4.4 $\pm$ 0.5 \\\noalign{\smallskip}
\hline
\end{tabular}
\end{table*}

The best simultaneous fit, shown overlaying all of the line profiles in Figure \ref{fig:lines}, gives the following central velocities and widths for each component. Ra: $V$ $=$ +120 $\pm$ 30 km s$^{-1}$, $\Delta V_{\rm FHWM}$ $=$ 350 $\pm$ 50 km s$^{-1}$; Rb: --130 $\pm$ 20, 230 $\pm$ 30 km s$^{-1}$;  B: --470 $\pm$ 10, 160 $\pm$ 10 km s$^{-1}$; VB: --720 $\pm$ 40, 370 $\pm$ 170 km s$^{-1}$. The full fit parameters are presented in Table \ref{tab:fits}. The advantage of the simultaneous fit is the robust constraint of VB and separation of Ra/Rb, even in spectra with the lowest signal-to-noise or coarsest velocity binning. The simultaneous fit is dominated by the high signal-to-noise [CII] spectrum. The final luminosity of each component in [CII] is $\la$1\% different from the best fit to that profile only.

We test whether [CII] is overly-dominant, simply forcing the `global' parameters to its own best fit, by repeating the fit with only the moderate signal-to-noise spectra: $^{12}$CO(2--1), $^{12}$CO(6--5) and [NII]. While the parameterization of the B and VB components in this test is essentially identical to the primary result, component Ra is fit by a larger Gaussian profile (and Rb by a correspondingly smaller Gaussian; Ra/Rb luminosity ratio changes from $\sim 1-2$ to $\sim 3-4$). In the full fit, we find that the Rb/B luminosity ratio is remarkably stable between lines ($\sim1.2\pm0.1$, excluding H$_2$O which is undetected for B), and is $>1$ for all the profiles. In the mid-SNR fit, Rb/B varies widely from $0.1-0.7$ (Rb is less luminous than B). Additionally, the red component fit to only mid-SNR lines is a lot less well constrained (twice the $\chi^2$). This test shows that although [CII] does dominate the solution, the fully-simultaneous method provides a more consistent description than individual, or even sub-set, fits.

There are four further issues to note about the best fit profiles shown in Figure \ref{fig:lines}. (1) The best simultaneous fit sets a 3$\sigma$ upper limit for the undetected VB component of $^{12}$CO(5--4) and $^{12}$CO(8--7): $I <0.9$ Jy km s$^{-1}$ ($L < 0.2\times10^9$ L$_\sun$) and $I <1.2$ Jy km s$^{-1}$ ($L < 0.9\times10^9$ L$_\sun$) respectively. (2) The [CI] and $^{12}$CO(7--6) profiles overlap, and consequently the VB component of the latter is buried in the red component of the former. The resulting luminosity for the $^{12}$CO(7--6) VB component is not well constrained, which also effects the luminosity of the B component and the [CI] Ra component. However, the best fit produces an Rb/B luminosity ratio of 1.1 for the $^{12}$CO(7--6) line, which is similar to the luminosity ratio in all the other lines, giving us confidence that the simultaneous fit has coped well even in this under-constrained case. (3) The H$_2$O emission line at 304 \micron{} is not detected for the B and VB components. We discuss the interpretation of this line later in the paper. (4) In C12, we interpreted the [NII] profile in terms of a wide, skewed blue peak. The simultaneous fit of a four component model provides a good fit to the [NII] profile, and suggests that the unusual blue peak actually results from a strong VB component. The relative luminosity of [CII] and [NII] in the Ra, Rb and B components is approximately equal, whereas VB shows significantly enhanced [NII]/[CII]. We explore this ratio further in Section \ref{sec:interp}.

\subsection{Spatially resolved emission}
\label{sec:maps}

We turn our attention to the spatially-resolved maps obtained using SMA and VLA, attempting to connect morphological structure to the spectral components identified in Section \ref{sec:lines}. The SMA 1mm continuum map (Figure \ref{fig:maps}) has the smallest beam, and reveals two bright components, separated by approximately 2$''$ and roughly orientated north/south. The brighter northern component particularly appears to be extended, while the two are linked by a low surface brightness bridge to the east. The flux ratio of the north/south components in the 1mm continuum is $\sim$1.7.

The SMA map of [CII] 158 \micron{} and the VLA map for the $^{12}$CO(1--0) transition both display a similar morphology to the 1mm continuum map (Figure \ref{fig:maps}), with bright north and south components clearly visible. However, both integrated line maps show a larger fraction of flux originating from between the north and south peaks, particularly to the west.

\begin{figure*}
\centering
\includegraphics[scale=0.29]{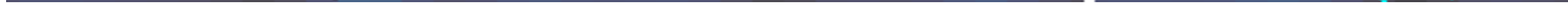}
\caption{[CII] ({\it upper row}) and $^{12}$CO(1--0) ({\it lower row}) maps showing $\sim10''\times10''$ region around HLS0918. North and east are aligned to the top and left, and the restoring beam size for each row is shown by a white ellipse in the lower-right corner. Each column displays a unique velocity slice corresponding to the spectral components identified in Section \ref{sec:lines}: --860 to --590 km s$^{-1}$ (VB), --590 to --310 km s$^{-1}$ (B), --310 to --30 km s$^{-1}$ (Rb), --30 to +420 km s$^{-1}$ (Ra). For each row, the colormaps are normalised to the peak intensity in the row, to highlight the relative intensity of each component. The black contours give the \{1,2,3...\} $\sigma$ confidence levels, while dashed cyan contours show the symmetric negative levels. Note that the relation between intensity and SNR for each panel is a function of the velocity bin width. The 1 mm continuum morphology from Figure \ref{fig:maps} is shown by white contours on each panel for reference. In [CII] emission, the spectral components correspond extremely well with distinct spatial entities. The division of the red peak into two components is vital to extract the western bridge. For $^{12}$CO(1-0), the spatial location of the Ra, B and faint VB components agree with [CII] within 1$\,\sigma$, given the beam sizes.}
\label{fig:comp_maps}
\end{figure*}

The [CII] and $^{12}$CO(1--0) datacubes have a high enough SNR to allow subdivision into velocity bins. We choose bins corresponding to the spectral components identified in Section \ref{sec:lines}. Specifically, bin boundaries are taken at velocities where overlapping components have equal contribution to the total flux (--860, --590, --310, --30, +420 km s$^{-1}$). Figure \ref{fig:comp_maps} displays the resulting maps. Concentrating first on the high signal-to-noise [CII] data (upper row), the spectral components are clearly located in distinct locations on the sky. Ra corresponds to the bright north and south peaks visible in the 1mm continuum map. The Rb and B appear as two arcs, to the west and east respectively, with the latter aligned to the bridge identified in the continuum map. Division of the broad `red' spectral peak into Ra and Rb components is essential to extract the western bridge. The faint VB spectral component appears as a faint source to the south-east of the continuum flux.

The $^{12}$CO(1--0) data has a lower signal-to-noise and a poorer spatial resolution than [CII], as highlighted by the significance ($\sigma$) contours in Figure \ref{fig:comp_maps}. Despite these limitations, we can identify many similarities to the [CII] maps. The spatial location of the two Ra peaks are within 1$\,\sigma$ (given the $^{12}$CO(1--0) beam size) of the corresponding [CII] peaks. Similarly, the $^{12}$CO(1--0) component B is dominated by the eastern bridge, and VB is entirely located to the south-east (albeit only at $\sim$3$\,\sigma$). In contrast, the Rb component is not in such good agreement, with the western bridge seen in the [CII] emission replaced by two peaks similar to the Ra component. One interpretation is that the [CII] and $^{12}$CO(1--0) emission in Rb originates from very different locations. However, the similarity of the location of Rb to Ra in the CO line instead suggests that the velocity width of the Ra component is larger in $^{12}$CO(1--0) than [CII], leaking into the adjacent bin and drowning out the Rb flux. Unfortunately, further investigation, such as constraining just the CO line profile fit with different parameter values, is hampered by the poor signal-to-noise of the VLA data, although we do note that the spectral peak for Ra in $^{12}$CO(1--0) is skewed towards lower velocities than in [CII] (see Figure \ref{fig:lines}). Such a difference in line width could be caused by the emission originating from different radii within the same SF cloud, but such an offset would be negligible compared to the beam size of our maps. Generally, the same structure appears in each velocity range for both [CII] and $^{12}$CO(1--0), verifying that they are not an artefact of instrumentation or unique to a particular line.

Finally, we estimate the apparent FIR luminosity of each component. We use the association between spectral and spatial components to derive the relative luminosity of each in the 1mm continuum map. Assuming that the 1mm continuum reflects the morphology of the continuum in general (and the shape of the dust SED in each component is comparable), we can then estimate $L_{\rm FIR}$ for each component. We approximate the north and south components in the 1mm map with a 2-d Gaussian profile each. Subtracting these from the map results in a residual image containing the spatially separated flux of the east and west bridges. We allocate flux to these pixel-by-pixel, using a line bisecting the gap. Two thirds of the continuum flux originates from the north and south components combined (spectral component Ra). The west and east regions (Rb and B respectively) contribute $\sim$10\% and 20\% of the total flux. We estimate the flux attributed to the VB component by assuming that the east component is a scaled version of the western source (VB is the residual from subtracting a rotated and scaled western component from the east). The VB component is attributed $\sim$1.5\% of the total flux. This allocation method accounts for $101\pm4$\% of the integrated continuum flux (first row, Table \ref{tab:derived}), which suggests that the continuum morphology is well modelled by the four components of the line emission.

The first row of Table \ref{tab:derived} presents the relative apparent continuum luminosities. $L_{\rm FIR}$ and SFR for each separate component are derived from these fractions, and are also shown in Table \ref{tab:derived}  (without correction for magnification).

\subsection{Lensing model}
\label{sec:lensing}

\begin{figure*}
\centering
\includegraphics[scale=0.7]{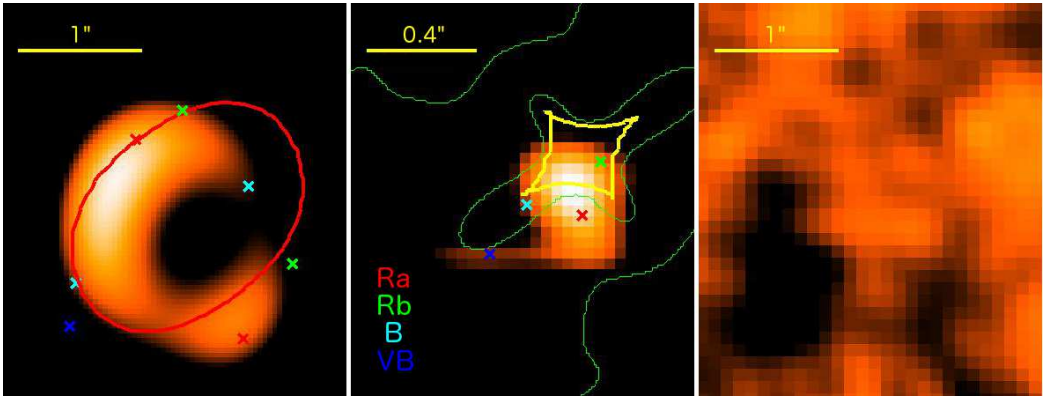}
\caption{The lensing model for HLS0918 (to be presented in detail in Boone et al., in prep.). {\it Left panel:} Sum of the flux in the four components for the model of HLS0918, after passing through the model lens and convolving with the SMA beam. The critical line is marked in red, while the crosses refer to the local maxima of each component in the image plane before convolution (i.e. directly through the lens). {\it Central panel:} Sum of the four components in the source plane (without SMA beam convolution). The caustic line is marked in yellow, while the crosses mark the position of each component peak. Magnifications of $\mu=2,5$ are shown by the green contours. $\mu=10$ lies very close to the caustic. {\it Right panel:} Residual map of total model flux in the image plane subtracted from the observation.}
\label{fig:lensmodel}
\end{figure*}

Derivation of the absolute luminosity, along with other source-plane characteristics such as the true morphology, requires a well-constrained model of the magnification and distortion due to lensing. For HLS0918, the lens model has to combine the effect of both Abell 773 ($z=0.22$) and the foreground galaxy at $z=0.63$. The large distance from the centre of A773 (5.6$'$) means that the cluster lens effect is secondary ($\sim10\%$ contribution). The best possible lens model is derived using {\sc LensTool} \citep{jul07-447}, via constraints obtained from both the SMA data described in this paper, and additional observations from the Plateau de Bure Interferometer (PdBI), to be presented in Boone et al. (in prep.). Comprehensive, quantitative details of the observed constraints and parameters for the lens model, including an in-depth analysis of differential lensing enabled by the higher spatial resolution PdBI imaging, will be given in Boone et al.

For the purposes of this paper, we only attempt to reconstruct the source-plane configuration of HLS0918 from the four components identified in the SMA [CII] and 1mm continuum maps. Figure \ref{fig:lensmodel} shows the derived source-plane configuration alongside the predicted SMA (image-plane) observation for HLS0918, given the best fit lens model, and the residual map of the model flux in the image plane subtracted from the observation. We find that the four components described in the previous section (plus two faint images of components Rb and B, which are not directly identified in the observed maps) originate from four source-plane regions separated by $\la4$ kpc (equivalent to $\la0.5''$ on the sky without a lens).

The bright north and south sources are indeed two images of the same region (together magnified by $\mu_{\rm Ra}=9.4\pm2.0$), as suggested by their identical velocity offset in the line profile. The vast majority of the source-plane flux is from Ra, which is located approximately half inside and half outside the caustic (although the peak is located outside; Figure \ref{fig:lensmodel}). In the image plane, the total flux is dominated by the part of component Ra which is crossing the caustic. The model indicates that both Rb and B also cross the caustic, producing double images with amplification factors of $\mu_{\rm Rb}=12.0\pm2.6$ and $\mu_{\rm B}=8.4\pm1.7$. In the observed SMA [CII] maps for the individual components (Figure \ref{fig:comp_maps}), the two images of Rb are not quite spatially resolved, while the western image of component B is only marginally visible: the Rb/Ra and B/Ra flux ratios are low, so the double image of Ra dominates in the frequency-integrated image-plane maps. The VB component originates from a region entirely outside of the caustic, and undergoes a lower magnification (although still $\mu_{\rm VB}=4.4\pm0.9$, as shown by the magnification contours in the central panel of Figure \ref{fig:lensmodel}) resulting in a single image. The total (`flux-weighted') magnification of the full source HLS0918 is $\mu_{\rm total}=8.9\pm1.9$. All of these magnification factors are listed in Table \ref{tab:derived}.

Combining the magnification factors with the apparent luminosities derived in the previous section, we compute the absolute FIR luminosity: $L_{\rm FIR,demag} = (1.8\pm0.4)\times10^{13}$ L$_\sun$. Of the components, Ra is by far the brightest with $L_{\rm FIR,Ra,demag} = (1.1\pm0.2)\times10^{13}$ L$_\sun$. Rb and B have similar luminosities to each other: $L_{\rm FIR,Rb,demag} = (1.9\pm0.4)\times 10^{12}$ and $L_{\rm FIR,B,demag} = (4.0\pm0.8)\times10^{12}$ L$_\sun$. The lower amplification of VB ($L_{\rm FIR,VB,demag} = (4.5\pm2.5)\times 10^{11}$ L$_\sun$) means that the source-plane luminosity ratio between that component and Rb or B is not as large as in the image plane.

Magnification is a strong function of source-plane position relative to the caustic line. If the emission regions are significantly more compact than the SMA observations can resolve, then these components may not extend as far beyond the caustic as currently assumed. We caution that the true magnification could be higher than estimated. The total magnification could be as large as $\mu_{\rm total}\sim15$. Similarly, the data presented in this paper does not allow us to constrain differential magnification, caused by size differences in the emission regions of different lines. We do, however, reiterate that the [CII] and $^{12}$CO(1--0) maps presented here are consistent (within 1$\,\sigma$ in flux and the VLA beam size) with a scenario in which all line emission from a particular component originates from a single region. A source-plane reconstruction utilizing all of the higher spatial resolution PdBI data may better constrain the size of the components, estimate the effect of differential magnification, and the total amplification (Boone et al., in prep.).

\section{Discussion}
\label{sec:discussion}

Emission line ratios provide a powerful probe of the physical conditions within the photon dominated regions of the ISM. Together with the lens model predictions of the source-plane configuration, we can characterize the nature of HLS0918 in great detail. We begin with the simple luminosity ratios $L_{\rm [CII]}/L_{\rm FIR}$ and $L_{\rm CO(1-0)}/L_{\rm FIR}$, which provide an easy way to connect various dust, gas and ISM properties in each component. To form a more complete characterization of the star-forming gas, we then move to the interpretation of multiple molecular and atomic lines via complex photo-dissociation region (PDR) models. We also re-examine the [CI], water and [NII] emission reported for HLS0918 in C12. Together, these three lines can help explore phenomenon vital to understanding the properties of HLS0918 in particular, but also the evolution of the high-redshift galaxy population in general: gas metallicity and feedback mechanisms via AGN and outflowing material.

Throughout this discussion, differential magnification is considered to have a negligible effect. Generally, this is a good approximation for bolometric fractions of [CII] and $^{12}$CO(1--0), and the current data provides no evidence for significant differences in spatial location of the components in those two lines (Figure \ref{fig:comp_maps}). However, the assumption is almost certainly poor for ratios within the $^{12}$CO SLED \citep{ser12-2429}, and hence we are careful to draw conclusions from order of magnitude estimates only. We facilitate comparison to previous studies by adopting the definition $L_{\rm FIR (42.5-122.5\mu m)}$ \citep[e.g.][]{sta10-957,bre11-8}, rather than $L_{\rm FIR (8-1000\mu m)}$, for both $L_{\rm [CII]}/L_{\rm FIR}$ and $L_{\rm CO(1-0)}/L_{\rm FIR}$, as noted in Table \ref{tab:derived}.

\subsection{$L_{\rm [CII]}/L_{\rm FIR}$}
\label{sec:cii_fir}

\begin{figure*}
\centering
\includegraphics[angle=270,scale=0.8]{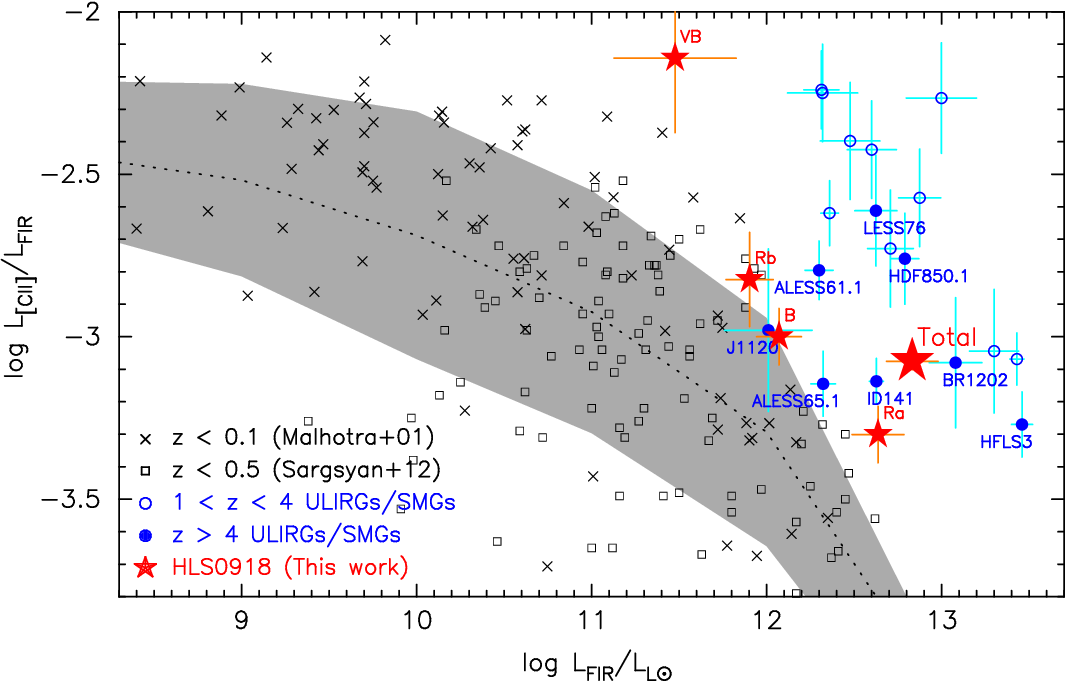}
\caption{$L_{\rm [CII]}/L_{\rm FIR}$ versus $L_{\rm FIR}$. HLS0918 (large red star) and constituent components (smaller red stars) assume magnifications as shown in Table \ref{tab:derived}. High redshift ($z>4$) ULIRGs/SMGs (filled blue circles; labeled with source name) from \citet{cox11-63}, \citet{bre11-8}, \citet{swi12-1066}, \citet{ven12-25}, \citet{wag12-30}, \citet{wal12-233}, \citet{rie13-329}. $1<z<4$ ULIRGs/SMGs (open blue circles) are from \citet{sta10-957} and \citet{val11-3473}.  Low redshift star-forming galaxies based on {\it ISO} data from \citet{mal01-766} and {\it Herschel}/PACS data from \citet{sar12-171} shown as crosses and open squares respectively. For the local data, we calculate the median (dotted line) and scatter (1$\,\sigma$ $=$ grey shaded area). HLS0918 is located in the same region as the other high luminosity, high redshift sources, which exhibit $L_{\rm [CII]}$ approximately an order of magnitude larger than local galaxies with similar FIR luminosity. The individual components of HLS0918 show an excess of 2--3.}
\label{fig:cii_fir}
\end{figure*}

$L_{\rm [CII]}/L_{\rm FIR}$ is a sensitive probe of physical conditions within the ISM. High-z sources often exhibit ratios an order of magnitude or more larger than local star-forming galaxies of similar FIR luminosity \citep[e.g.][]{sta10-957,swi12-1066}, which is usually attributed to large, extended [CII] reservoirs.

Figure \ref{fig:cii_fir} displays this ratio as a function of FIR luminosity, correcting for the effect of magnification. The integrated source has $L_{\rm [CII]}/L_{\rm FIR}$ $=$ (8.4 $\pm$ 0.5) $\times$10$^{-4}$, which lies significantly above nearby ULIRGs and is consistent with other star-forming galaxies at $z>4$. Generally, these galaxies exhibit a smaller offset from the local ULIRG population than those at the epoch of peak star formation ($z\sim2$), although selection bias and small number statistics are unaccounted for. Component Ra exhibits a lower ratio [(5 $\pm$ 1) $\times$10$^{-4}$], while components Rb and B are similar to each other [(1.5 $\pm$ 0.5) and (1.0 $\pm$ 0.2) $\times$10$^{-3}$] and larger than Ra. Individually, components Ra, Rb and B are much closer to the locus of local star-forming galaxies than the integrated source: individual high-z clumps may not be as different from local star-formation as the integrated galaxies suggest. VB has stronger [CII] emission given its FIR luminosity [(7 $\pm$ 4) $\times$10$^{-3}$], which is more similar to much lower FIR luminosity systems in the local Universe. This may indicate that VB lacks an extended [CII] reservoir.

\subsection{Gas mass and star formation efficiency ($L_{\rm CO(1-0)}/L_{\rm FIR}$)}
\label{sec:gasmass}

The stability and abundance of CO molecules in interstellar clouds (predominantly composed of H$_2$), implies that the ubiquitous emission from CO rotational transitions is an important tracer of total gas mass. As $L_{\rm FIR}$ quantifies star formation in dusty systems, the $L_{\rm CO(1-0)}/L_{\rm FIR}$ ratio is therefore a useful diagnostic for the star formation efficiency (SFE). The VLA $^{12}$CO(1--0) observation of HLS0918 is particularly important, removing the need to extrapolate $L_{\rm CO(1-0)}$ from higher-$J$ transitions.

For the integrated source $L_{\rm CO(1-0)}/L_{\rm FIR}$ $=$ (4.3 $\pm$ 0.4) $\times$ 10$^{-7}$, which sits towards the upper-envelope (i.e. most efficient) of $z\sim2-3$ literature values \citep[e.g.][]{bre11-8}, but is similar to the ratio for the $z=2.3$ `Eyelash' galaxy (SMMJ2135) reported to have $L_{\rm CO(1-0)}/L_{\rm FIR}$ $=$ 5$\times$10$^{-7}$ \citep{dan11-1687}. SMGs at $z>5$ show a similar spread in this ratio as their $z\sim2-3$ counterparts, e.g. HDF850.1 \citep[$z=5.2$,][]{wal12-233}, HFLS3 \citep[$z=6.3$,][]{rie13-329} with $L_{\rm CO(1-0)}/L_{\rm FIR}$ $\sim$ $1-2 \times$ 10$^{-7}$. Components Ra and B exhibit comparable $L_{\rm CO(1-0)}/L_{\rm FIR}$ to each other, resembling more the higher SFEs of intermediate redshift ($z\sim1-2$) ULIRGs \citep[e.g.][]{com11-124,com13-41}. Component VB has an extremely elevated $L_{\rm CO(1-0)}/L_{\rm FIR}$, although we note the large error bar that suggests the component is on the limit of detection in the $^{12}$CO(1--0) profile.

We quantify the total gas mass from the lensing-corrected $^{12}$CO line luminosity\footnote{CO line luminosity (in K km s$^{-1}$ pc$^2$) via
$L'_{\rm CO(1-0)} = 3.25\times10^7 \cdotp I_{\rm CO} \cdotp \nu_{\rm obs}^{-2} \cdotp D_{\rm L}^2 \cdotp (1+z)^{-3}$
with integrated luminosity $I_{\rm CO}$ in Jy km s$^{-1}$, $\nu_{\rm obs}$ in GHz and luminosity distance $D_{\rm L}$ in Mpc.} via the basic relation
\begin{equation}
M_{\rm H_2} = \alpha_{\rm CO} L'_{\rm CO(1-0),unlensed}
\end{equation}
where $M_{\rm H_2}$ ($M_\sun$) is defined to include helium such that $M_{\rm H_2}$$=$$M_{\rm gas}$ \citep{sol05-677} and $\alpha_{\rm CO}$ [$M_\sun$ (K km s$^{-1}$ pc$^2$)$^{-1}$] is the conversion factor. Assuming that the $^{12}$CO(1--0) emission is optically thin and the environment is solar metallicity, we can estimate a lower limit for $\alpha_{\rm CO}$, and hence total gas mass. Following \citet{ivi10-198} and \citet{dan11-1687}
\begin{multline}
\alpha_{\rm CO} = 0.08 \left\{ \frac{g_1}{Z} e^{-T_0/T_k} \left( \frac{J(T_k) - J(T^{z}_{\rm CMB})}{J(T_k)} \right) \right\} ^{-1} \\ \times \left( \frac{[^{12}{\rm CO}/{\rm H_2}]}{10^{-4}} \right) ^{-1}
\end{multline}
where $g_1$ $=$ 3 (the degeneracy of level $n=1$), the partition function $Z\sim2(T_k/T_0)$, $T_0$ $=$ 5.5 K, $J(T)=T_0(e^{T_0/T}-1)^{-1}$, and $[^{12}{\rm CO}/{\rm H_2}]$ $\sim$ $10^{-4}$ \citep[based on Mk231,][]{bry96-678}. We adopt a lower limit on gas temperature $T_k$ $>$ 40 K, as gas couples well to warm dust in a dense star-forming system \citep{nar11-664}. For HLS0918 at $z$$=$5.2430, $T^{z}_{\rm CMB}$ $=$ (1+$z$)$T^{z=0}_{\rm CMB}$ $\sim$ 17 K, and we therefore estimate $\alpha_{\rm CO}$ $\ga$ 0.7, which is consistent with the value for a smoothly distributed, largely molecular ISM, as observed in local ULIRGs ($\alpha_{\rm CO}$ $\sim$ 0.8; e.g. \citealt{sol05-677}).

The lower limit on the total gas mass for HLS0918 is $M_{\rm gas}$ $\ga$ (74 $\pm$ 17) $\times$ 10$^{9}$ $M_\sun$. Assuming that $\alpha_{\rm CO}$ is the same for all components, Ra contains almost half the mass, $M_{\rm gas}$ $\ga$ (31 $\pm$ 10) $\times$ 10$^{9}$ $M_\sun$, while the remainder is distributed roughly equally between Rb, B and VB.

We can now use the absolute gas mass to calculate the SFE in physical units: SFE (in L$_\sun$ M$_\sun^{-1}$) $=$ $L_{\rm FIR}/M_{\rm gas}$ \citep{gre05-1165}. HLS0918 has a maximum SFE $\la$250 $\pm$ 20 L$_\sun$ M$_\sun^{-1}$, which does not conflict with the absolute limit due to radiation pressure derived by \citet{sco04-253}, SFE$_{\rm max}$ = 500 L$_\sun$ M$_\sun^{-1}$. Individually, components Ra, Rb and B have SFE in the range 130--360 L$_\sun$ M$_\sun^{-1}$. Component VB exhibits a lower SFE $\la$60 $\pm$ 60 L$_\sun$ M$_\sun^{-1}$. Although the apparent continuum flux of VB is significantly lower than the other components, the total molecular gas mass is similar to Rb and B. This is due to both a smaller amplification effectively decreasing the overall relative luminosity of VB, and an intrinsically lower SFE which results in an enhanced $L_{\rm CO(1-0)}/L_{\rm FIR}$. HLS0918 corresponds well with the general population of high-redshift SMGs, averaging $\sim$260 L$_\sun$ M$_\sun^{-1}$ \citep[e.g.][]{rie10-153,dan11-1687,tho12-2203}.

\begin{figure*}
\centering
\includegraphics[angle=270,scale=0.8]{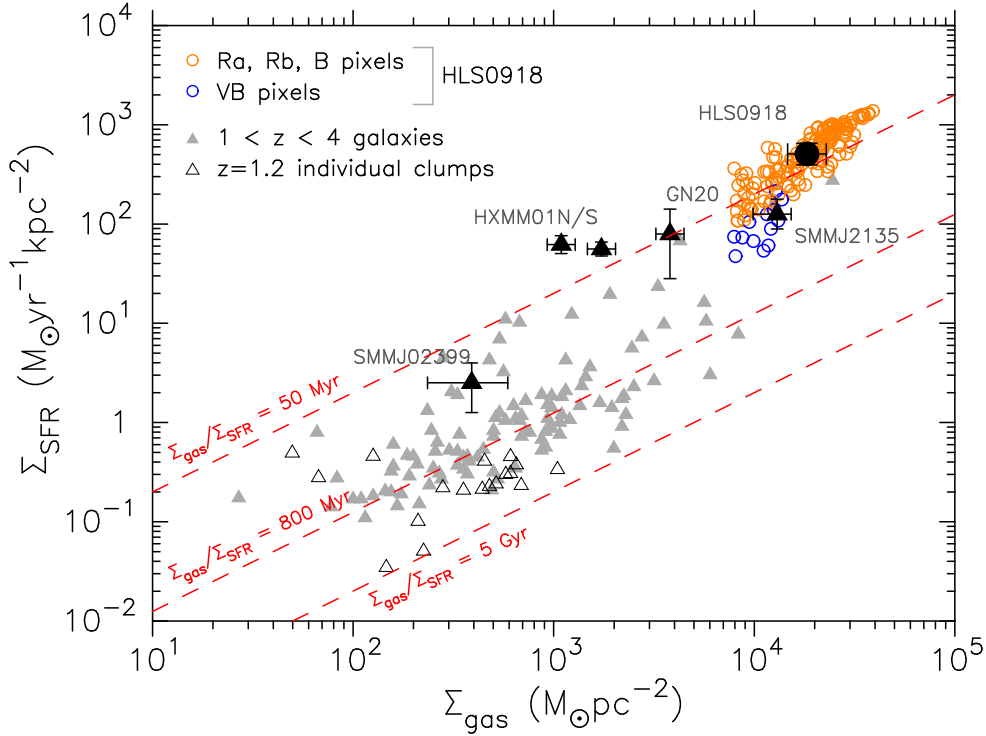}
\caption{Kennicutt--Schmidt relation: SFR and total molecular gas surface densities ($\Sigma_{\rm SFR}$ vs $\Sigma_{\rm gas}$). The mean surface densities for HLS0918 are shown by the black circle. Open circles display the per-pixel values: pixels dominated by flux from component VB (highlighted in blue) are located beyond the general locus of values for HLS0918.  Note that the VB pixels are all correlated in $\Sigma_{\rm SFR}$, as the component is not resolved by the $^{12}$CO(1--0) beam. Filled triangle show additional star-forming galaxies ($z\sim1-4$; \citealt{gen10-2091,tac13-74}), with well-known bright sources labelled \citep{car10-1407,ivi10-198,dan11-1687,fu13-338}. Open triangles display individual star-forming clumps within four massive galaxies at $z=1.2$ \citep{fre13-130}. The red dashed lines indicate constant gas consumption timescales for the star-forming regions of 50 Myr, 800 Myr and 5 Gyr. The $1<z<4$ star-forming population exhibit a mean of $\sim$800 Myr, while HLS0918 lies above the upper extreme with a consumption timescale $<50$ Myr.}
\label{fig:resolved_ks}
\end{figure*}

Finally, we examine the Kennicutt--Schmidt (KS) relation, a power law linking the surface densities of star formation rate and molecular gas ($\Sigma_{\rm SFR}$ and $\Sigma_{\rm gas}$). The trend is near linear for a wide range of gas densities ($\Sigma_{\rm gas} \ga 0.1$ M$_\sun$ pc$^{-2}$) and at all observed redshifts \citep[e.g.][]{ken98-189,gen10-2091,dec12-2,tac13-74}, suggesting that the mode of star formation is relatively consistent throughout time. We estimate the mean surface densities for HLS0918 from the resolved emission in the SMA 1mm continuum and VLA $^{12}$CO(1--0) maps, which can be converted to SFR and gas mass respectively, using the prescriptions described in earlier sections. HLS0918 is located at the upper envelope of the general KS relation (Figure \ref{fig:resolved_ks}), along with other bright, high-z sources such as GN20 at $z=4.05$ \citep{car10-1407}. $\Sigma_{\rm gas}$/$\Sigma_{\rm SFR}$ is a crude estimate of the consumption timescale, with the estimate for the $1<z<4$ star-forming population located between $\sim$50 Myr and 5 Gyr, with a mean $\sim$800 Myr. HLS0918, along with GN20 and HXMM01 \citep[][$z=2.3$]{fu13-338}, all show shorter gas consumption timescales than this (10--50 Myr). However, $\Sigma_{\rm gas}$ is derived from $^{12}$CO($J>1$) for the $1<z<4$ galaxies in Figure \ref{fig:resolved_ks}, which may overestimate the density compared to $^{12}$CO(1--0). The paucity of lower redshift galaxies with either $^{12}$CO(1--0) observations, or such large SFR surface densities, means that a direct comparison is difficult: are these bright examples offset due to an intrinsic difference in the mode of star formation, or is this a selection effect whereby we are only probing the upper edge of the population?

We can examine the resolved KS relation for HLS0918. Although we can separate the individual components, we do not know the spatial extent of each clump and cannot directly determine the surface densities. Instead, we grid the spatial data to calculate the mean surface densities for different regions of HLS0918 (in the image plane). We adopt the $^{12}$CO(1--0) map pixel size ($0.3\times0.3''$) as the most convenient binning regime. The continuum map is smoothed to the $^{12}$CO(1--0) beam and re-binned, to give SFR surface density in bins exactly matched to the gas surface densities. The densities are corrected for mean component amplifications, but we do not account for pixel-scale magnification gradients. Gravitational lensing conserves surface brightness Figure \ref{fig:resolved_ks} shows that the binned points form a generally elliptical locus about the mean value. A small fraction of pixels are offset from the main population (at lower SFR surface density for a given $\Sigma_{\rm gas}$). Although it is hard to distinguish the effect of components Rb and B from the dominant Ra, component VB is more isolated spatially. We find that the offset pixels are from the region coincident to VB, presenting further visual evidence that VB has an intrinsically lower SFE than the rest of HLS0918. However, VB is certainly not atypical compared to other star-forming sources, with the bright lensed galaxy SMMJ2135 \citep{dan11-1687} located in the same region of the KS plot.

\subsection{Physical properties of the PDR}
\label{sec:pdr}

For a more detailed exploration of the ISM properties, we now employ the solar metallicity PDR models of \citet[][hereafter K99]{kau99-795} and the higher metallicity (twice solar) models from \citet[][M07]{mei07-793}. Both models parameterize PDR characteristics in terms of atomic gas density ($n$ in cm$^{-3}$) and incident FUV flux from young massive stars ($G_0$ in Habing fields). K99 present a simple model where the assumed geometry is of a cloud illuminated on one side. A more realistic model would assume illumination of the cloud on all sides, and therefore optically thick emission is detected only from the near-side of the cloud and optically thin emission is detected from both near and far sides, affecting the observed line ratios. Corrections may be applied to account for this, but we choose not to here. Instead, we also compare to a more geometrically complicated model from M07 which involves two or more clouds with differing densities and incident radiation fields. In both models we assume that all the observed emission lines in a single component are coming from the same gas, which is a vast oversimplification, and again, we do not account for differential magnification. We do, however, correct the observed line luminosities to account for the contribution from gas excitation by the warmer CMB at $z>5$ ($T^{z=5.24}_{\rm CMB}$ $=$ 17 K). The appropriate correction for each line is estimated via the {\sc radex} code \citep{tak07-627}, assuming $T_k$ $=$ 40 K. Corrections are small, but not negligible: averaged over likely gas densities ($\log n\sim2-6$) we estimate $\sim40$\% for $^{12}$CO(1--0), 15\% for high-$J$ CO transitions, and 4\% for [CII], which agrees well with \citet[][Figure 10]{dac13-13}. We discuss these corrections in more detail later in this section. Overall, the simplifications in modelling restrict our discussion to order-of-magnitude estimates of the characteristic properties.

\begin{figure*}
\centering
\includegraphics[angle=270,scale=0.9]{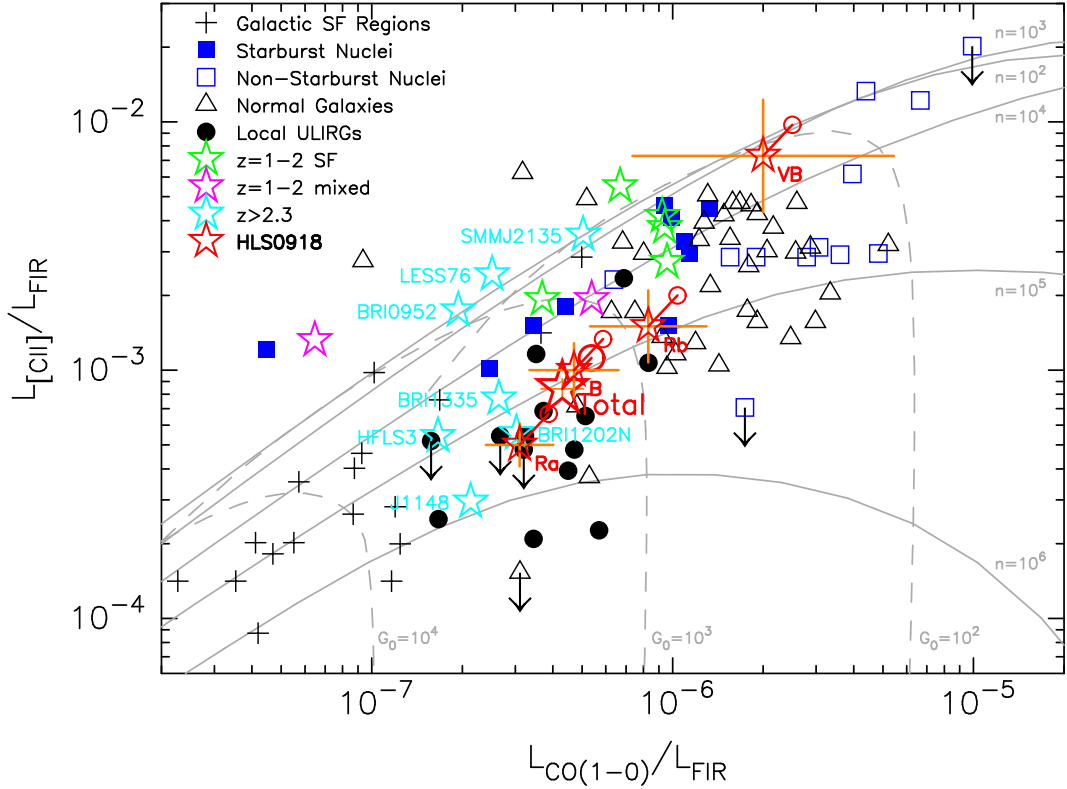}
\caption{$L_{\rm [CII]}/L_{\rm FIR}$ versus $L_{\rm CO(1-0)}/L_{\rm FIR}$ diagnostic plot (adapted from \citealt{bre11-8}). HLS0918 (labeled red stars; integrated and spectral components plotted separately) is compared to a variety of local and high redshift sources. For consistency with the high-redshift sources taken from the literature, we plot the HLS0918 ratios un-corrected for CMB heating effects (labelled red stars). The grey lines represent the solar metallicity PDR models with varying gas density ($n$) and FUV field strength ($G_0$) from \citet{kau99-795}. However, these models are for the CMB temperature at $z=0$, and we note that at $z>4$ the effect of the warmer CMB becomes increasingly more significant. Thus for HLS0918, we also artificially correct for $T^{z=5.24}_{\rm CMB}$ $=$ 17~K, which increases the ratios by $\sim$0.2 dex (open red circles, linked by red lines to the corresponding un-corrected point) and systematically moves the points towards lower $n$ and $G_0$  (see text for details). Regardless of these corrections, components Ra, Rb and B appear similar to local ULIRGs, and Ra also most resembles high-redshift quasar host galaxies and the high redshift SMG HFLS3 \citep{rie13-329}. VB exhibits the physical characteristics of a normal (or even non-starbursting) galaxy.}
\label{fig:ngplot}
\end{figure*}

We begin with the simplest analysis: direct comparison of $L_{\rm [CII]}$ / $L_{\rm FIR}$ and $L_{\rm CO(1-0)}$ / $L_{\rm FIR}$ \citep{sta91-423,sta10-957,bre11-8}. Figure \ref{fig:ngplot} presents the location of HLS0918 and the constituent components on this diagnostic plot, with typical PDR characteristics indicated by a simple inversion of the K99 model. These contours assume that the [CII], $^{12}$CO(1--0) and FIR emission is PDR-driven, as the CMB heating effect at $z=0$ (where $T^{z=0}_{\rm CMB}$ $=$ 2.73 K) is negligible. For HLS0918, we plot both the observed ratios, and those including the correction for CMB heating at $z=5.2$, as described above ($\sim$0.2 dex shift). The correction tends towards both lower gas densities and lower $G_0$. The corrections and uncertainties ensure that a direct estimate of $n$ and $G_0$ is not meaningful, but we can compare the general location of the HLS0918 components to other observed galaxy populations. The integrated emission from HLS0918 is similar to local ULIRGs, and less than 0.5 dex from three well studied high-redshift sources: the SMG HFLS3 ($z=6.34$; \citealt{rie13-329}), the quasar host galaxy BRI1335--0417 ($z=4.41$; \citealt{rie08-9}), and BRI1202--0725N, the SMG component of a quasar/SMG merger at $z=4.7$, which cannot be deblended from its companion at some wavelengths (\citealt{mom05-1809,sal12-57}). In contrast, two other well-studied high-z SMGs, SMMJ2135 ($z=2.3$; \citealt{swi10-733,ivi10-35}) and LESSJ0332 ($z=4.76$; \citealt{bre11-8}) are more than half a magnitude larger in $L_{\rm [CII]}$ / $L_{\rm FIR}$, which is interpreted as an atomic gas density more than a magnitude lower than HLS0918.

Component Ra exhibits the highest gas density and strongest incident FUV field, which may indicate a very intense nuclear starburst or the presence of an AGN. The latter explanation is perhaps supported by the similarity of Ra in these luminosity ratios to two known, high-z quasar hosts. Figure \ref{fig:ngplot} shows that components Rb and B have lower densities and generally weaker FUV radiation fields than Ra, suggesting a mode of star formation more analogous to local LIRGs: i.e. star formation occurring on larger physical scales rather than in a compact region. This suggests that the star formation prescriptions derived from the brightest galaxies in the local Universe may not be applicable to the extreme star-forming SMGs, despite both exhibiting similar (ULIRG) luminosities. Finally, the luminosity ratios of VB are not as well constrained, but indicate that the component is much lower density and less vigorous in star formation. The wide dispersion of individual components within HLS0918 in Figure \ref{fig:ngplot} highlights the general difficulty in interpreting the location of composite (integrated) sources in such a plot.

\begin{figure*}
\centering
\includegraphics[trim=0mm 0mm 35mm 0mm,clip,angle=270,scale=0.62]{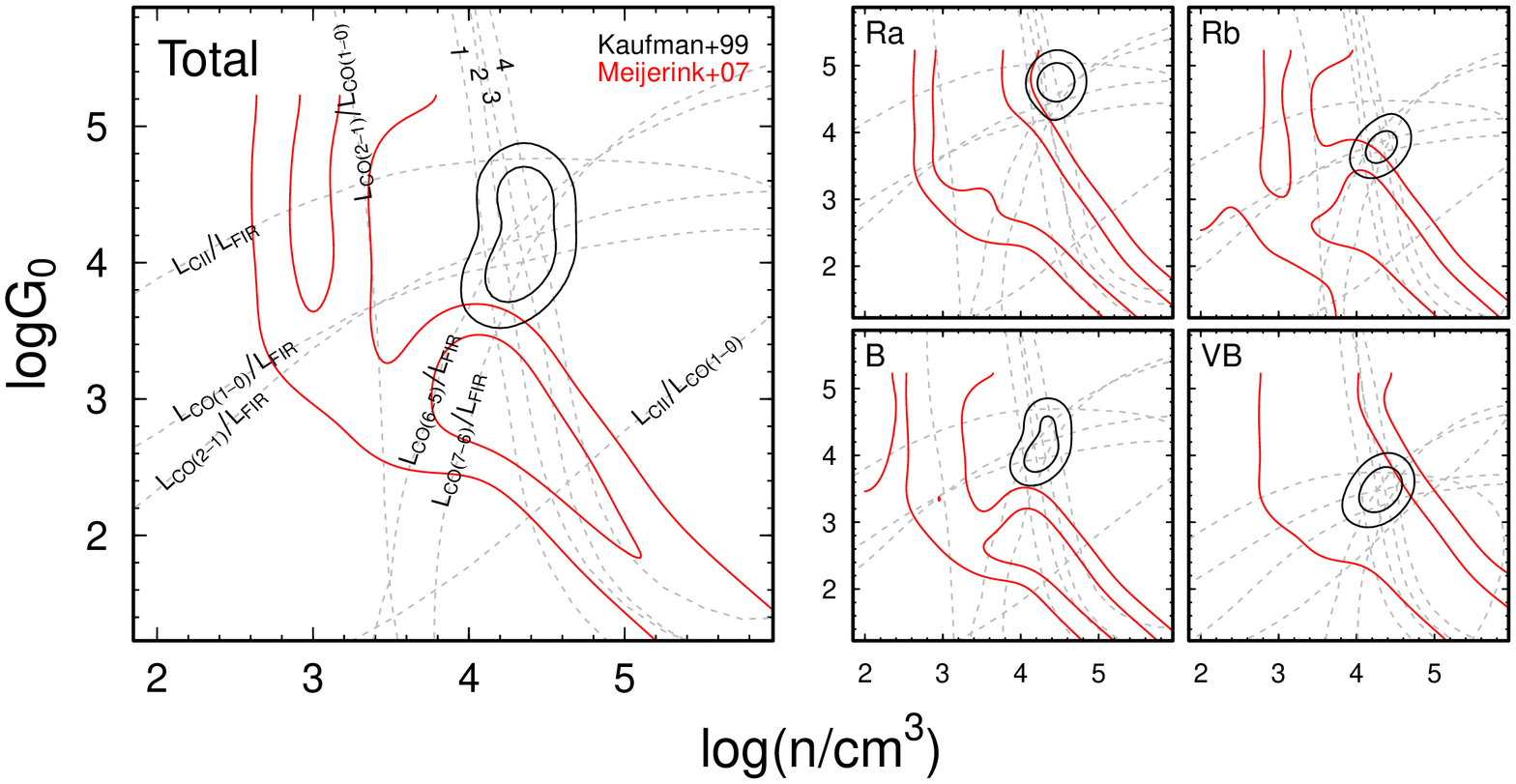}
\caption{HLS0918 luminosity ratios involving $^{12}$CO, [CII] and $L_{\rm FIR (42.5-122.5\mu m)}$ (dashed lines) as a function of gas density ($n$) and incident FUV flux ($G_0$, in units of Habing field), from the PDR models of \citet{kau99-795}. The observed ratios have been adjusted to account for the non-negligible CMB heating at $z=5.24$. The correction factors may be underestimated at the lowest gas densities (see text for further details). For clarity the ratios $L_{\rm CO(6-5)}$/$L_{\rm CO(1-0)}$, $L_{\rm CO(7-6)}$/$L_{\rm CO(1-0)}$, $L_{\rm CO(6-5)}$/$L_{\rm CO(2-1)}$ and $L_{\rm CO(7-6)}$/$L_{\rm CO(2-1)}$ are labeled 1, 2, 3 and 4 respectively. The black solid contours show the peak likelihood solution for the \citeauthor{kau99-795} models, while the red contours show similar contours from the less well constrained \citet{mei07-793} PDR models (including only [CII] and $^{12}$CO transitions). The main (left) panel presents the integrated system, while on the right side, the four individual components are shown. Ra, Rb and B exhibit very similar PDR diagnostic plots to the total system, exhibiting characteristics suggestive of extended $^{12}$CO(1--0) and [CII] gas. The diagnostic diagram of the VB component is striking for a possible single-phase solution.}
\label{fig:pdr}
\end{figure*}

We now explore the full complement of observed lines, whilst remembering that differential magnification will distort luminosity ratios involving both low- and high-$J$ $^{12}$CO lines. Figure \ref{fig:pdr} presents the individual luminosity ratios involving [CII], the $J_{\rm upper}=(1,2,6,7)$ $^{12}$CO transitions, and $L_{\rm FIR}$
as a function of $n$ and $G_0$ (K99 model). We convert the allowed parameter space for each line ratio into a probability distribution and derive the likelihood of each $n-G_0$ combination, also displayed in Figure \ref{fig:pdr}. The maximum likelihood from the K99 model solution is well constrained. In contrast, the M07 models consider a more complex geometry than K99, and are specifically tailored for systems with concentrated emission from galaxy centers and AGN, but do not predict $L_{\rm FIR}$. Furthermore, interpretation using the M07 models is impeded by a rather degenerate maximum likelihood solution, dominated by high-$J$ ratios with severely non-linear tracks in $n-G_0$ space.

We concentrate on interpreting the K99 model contours in Figure \ref{fig:pdr}. We remember that the K99 models cannot be applied directly at high redshift, where there is a warmer CMB temperature. For each line ratio, we have applied the corrections introduced above. However, these corrections are medians  over the full range of likely gas densities (the plotted range in Figure \ref{fig:pdr}). In reality, the corrections for CMB heating are a function of the gas density itself, and should be significantly larger at lower gas densities. To test the validity of the median corrections, we return to the {\sc radex} code and this time calculate the maximum likelihood gas density, given the observed line ratios and once again assuming $T_k$ $=$ 40 K. We find that the the gas densities are systematically half a dex lower than estimated via the PDR model (for the integrated source, $n=10^{3.9\pm0.2}$ cm$^{-3}$ compared to $n\approx10^{4.3}$ cm$^{-3}$ from Figure \ref{fig:pdr}), but within the $3\,\sigma$ contours. However, the {\sc radex} method includes only line ratios and ignores $L_{\rm FIR}$ as a constraint, as is the case for the M07 PDR models. In Figure \ref{fig:pdr}, the M07 solution also favors a much lower gas density than K99, so we do not know to what extent the lower gas density from {\sc radex} is due to CMB heating. In the following qualitative discussion, we consider the K99 PDR solution, and any conclusions are robust against the possible $\sim$0.5 dex overestimation of $n$.

Components Ra, Rb and B appear to exhibit very similar PDR characteristics to each other. As both critical density and excitation energy increases with increasing $J$, higher-$J$ $^{12}$CO transitions trace the denser regions, and ratios involving their luminosity are naturally located towards higher $n$ in the diagnostic plot. The $L_{\rm CO(2-1)}$/$L_{\rm CO(1-0)}$ contour is parallel to these, but at lower $n$, which suggests that $^{12}$CO(1--0) is the most extended. From observations of three quasar host galaxies at $z\sim4$, \citet{rie06-604} estimated that an extended molecular gas component could contribute up to 30\% of the overall emission line luminosity of $^{12}$CO(1--0). Unfortunately, the spatial resolution of the $^{12}$CO(1--0) map for HLS0918 hampers the direct detection of extended $^{12}$CO emission, showing no significant structure beyond the SMA 1mm continuum map convolved by the VLA restoring beam. However, if we assume that extended emission accounts for 30\% of the observed flux, the densest PDR regions (i.e. 70--75\% of $^{12}$CO(1--0); 100\% of higher-$J$ lines) presents a single-phase solution, suggesting a compact, uniform star-forming nucleus.

The other interesting discrepancy is the $L_{\rm [CII]}$ / $L_{\rm CO(1-0)}$ contour, which lies at significantly lower $G_0$ than the peak likelihood solution. This suggests that there may also be [CII] emission arising from a low density, cold, neutral ISM component, as proposed by e.g. \citet{mad93-579}, who observed as much as 50\% of [CII] originating from beyond the dense PDR. HLS0918, as well as many other high redshift galaxies, exhibits a larger $L_{\rm [CII]}$ / $L_{\rm FIR}$ for a given FIR luminosity than the local star-forming galaxies (Figure \ref{fig:cii_fir}). An extended [CII] reservoir surrounded star-forming regions at high redshift, but lacking from the more evolved local galaxies (possibly through stripping or quenching processes), would account for such an offset. Indeed, remove 50\% from the observed $L_{\rm [CII]}$ and components Ra, Rb and B would lie very close to the local galaxy population mean in Figure \ref{fig:cii_fir}.


The VB component exhibits a marginally lower gas density and $G_0$ compared to the other components. Furthermore, there is a unique intersection of all the luminosity ratios included in the K99 models. If the multi-phase solution for Ra, Rb and B are due to extended [CII] and $^{12}$CO(1--0) gas, then by the same logic the VB component PDR diagram may be interpreted as evidence for a lack of such an extended gas component. The lack of significant detections for VB in the high-$J$ $^{12}$CO lines also indicates that the gas in this component is not just at lower density, but is also cooler. We tentatively interpret this as evidence for a much less vigorously star-forming region than the intense R and Rb (also revealed by a lower SFE), and suggest that the galaxy hosting component VB may have undergone a process whereby extended gas has been removed (e.g. stripping via interaction). However, we caution that all of the measured $^{12}$CO emission in the VB component is near the limit of detection.



\subsection{Interpretation of H$_2$O, [NII] and [CI]}
\label{sec:interp}

The PDR analysis based on the [CII] and $^{12}$CO emission tends towards a scenario in which the four components originate from the star-forming regions in a system of 2--3 interacting galaxies. Additional line detections shed more light on this configuration, and in the case of component VB, may provide evidence for an alternative, exotic origin.

The H$_2$O line profile at 304 \micron{} is particularly interesting for only being detected in components Ra and Rb (and not B and VB). Although water is the second most abundant oxygen-bearing molecule in the warm ISM after CO, a very high infrared radiation field is required to liberate ice through grain heating, and to excite detectable emission lines. Recently, \citet{omo13-115} reported a ubiquitous power law $L_{\rm H_2O}=L_{\rm FIR}^\alpha$ for ULIRGs at all redshifts, where $\alpha=1.22\pm0.10$ for H$_2$O$_p$(2,0,2-1,1,1). Ra is consistent with this trend, exhibiting $\alpha=1.13\pm0.09$. In contrast, Rb has very strong water emission relative to $L_{\rm FIR}$ ($\alpha=2.5\pm0.7$), with a comparable intrinsic FIR luminosity to Apr 220, but 3--4 times the $L_{\rm H_2O}$. None of the ULIRGS presented in \citeauthor{omo13-115} exhibits such a large H$_2$O/FIR ratio, although a direct comparison is impossible as the sample does not include high-redshift galaxies with such a large FIR luminosity. The most likely scenario is that the strong FIR radiation from intense star formation\footnote{An embedded AGN might also produce the required radiation field to excite the water emission. Figure \ref{fig:ngplot} indicates a resemblance between Ra and high-redshift quasar hosts, so this possibility cannot be ruled out on the current evidence. Further constraint on a possible AGN is difficult without mid-IR or optical emission line detections.} in component Ra ($L_{\rm FIR,Ra}\approx 10^{13}$ L$_\sun$ after accounting for lensing) excites water emission in both Ra and Rb, which in turn indicates that the two components are neighboring star-forming regions. The lens model indicates a source-plane separation between Ra and Rb of  $<$1 kpc, and the components have a mutual $\Delta V$=250 km s$^{-1}$, which is consistent with two regions within the nucleus of a massive galaxy.

Comparing components B and Rb, water emission is the most discrepant line. In every other line profile B and Rb exhibit consistent ratios of Rb/B $=1.2\pm0.1$ (Figure \ref{fig:lines}). With the H$_2$O emission in Rb now explained via excitation through close proximity to the intense FIR radiation from Ra (which is not the case for B), the similarity in all other respects is striking. We speculate that the star formation in Rb and B were triggered by the same event, resulting in similar timescales and hence consistent physical gas properties. It is possible that B is a spiral arm, or tidal tail on the outskirts of the Ra--Rb system. However, with such a large velocity offset from the `nuclear' Ra component ($\Delta V \sim 600$ km s$^{-1}$), it may be more likely that component B is a region of triggered nuclear star formation in a second galaxy, interacting with Ra--Rb ($\sim$1:3 major merger, based on continuum luminosity and derived gas masses). The lack of detected water emission in component B suggests that the star formation is much lower density than the nuclear Ra (e.g. M82, \citealt{duf87-68}). The merger scenario is also consistent with recent studies \citep[e.g.][]{eng10-233} suggesting that the vast majority of SMGs are gas-rich major mergers (mass ratios of 1:3 or greater), with an increasingly large number of well constrained examples \citep[e.g.][]{tac08-246,dan11-1687,agu13-164,fu13-338}. The fraction of late-stage merging systems within the local ULIRG population ($\sim$64\% have nuclear separations $\le$ 2 kpc; \citealt{vei02-315}) indicates that many (unresolved) compact SMGs are recently coalesced. HLS0918, although resolved due to the strong lensing effect, falls into this category, as components Ra and B have a line-of-sight, source-plane separation of only $\sim$1 kpc, 

The $L_{\rm [NII]}/L_{\rm [CII]}$ ratio is an indicator of ionization and potentially traces metallicity. In the local Universe ($z \ll 0.1$), [NII]/[CII] is reported for only six galaxies (mean $\sim$ 0.05, ranging from 0.01--0.07), with {\it Herschel} spectroscopy providing many recent observations: M82, Mrk231, NGC1097, Arp220, NGC1068 and NGC4559 \citep{pet94-17,fis10-41,bei10-60,ran11-94,spi12-108,cro12-81}. ALMA will facilitate a similar increase in data at high redshift in the near future, but thus far, only one high-redshift $L_{\rm [NII]}/L_{\rm [CII]}$ measurement has been reported, for LESSJ0332 at $z=4.76$ \citep{nag12-34}. With the new [CII] observation from SMA and [NII] reported by C12, HLS0918 is the second high-z source with both line measurements.

The integrated HLS0918 has $L_{\rm [NII]}/L_{\rm [CII]}$ $=$ 0.052 $\pm$ 0.006, which is coincidentally identical to $L_{\rm [NII]}/L_{\rm [CII]}$ $=$ 0.052 $\pm$ 0.006 reported for LESSJ0332. However, it is the de-convoluted line profile of [NII] which is particularly interesting, with a striking difference compared to the [CII] and $^{12}$CO lines. Whereas $L_{\rm [NII]}/L_{\rm [CII]}$ is consistent within the errors for Ra, Rb and B (0.039 $\pm$ 0.016, 0.047 $\pm$ 0.023 and 0.039 $\pm$ 0.017 respectively), component VB has a larger ratio: $L_{\rm [NII]}/L_{\rm [CII]}$ $=$ 0.12 $\pm$ 0.07. Even though the formal uncertainty from the fit is high, the luminosity ratio of VB is visibly larger in the plotted profiles (Figure \ref{fig:lines}), due to the exceptionally bright [NII] emission.

An enhanced [NII]/[CII] may indicate a high metallicity. From the models presented in \citet{nag12-34}, we estimate a gas metallicity for VB of $\sim$2--3 $Z$/$Z_{\rm \sun}$, whereas the other components appear approximately solar metallicity. Although high redshift galaxies are generally expected to be less enriched than local counterparts, an intensely active system such as HLS0918 may have significantly enriched gas, although it is not clear why this should strengthen the [NII] emission of only component VB, which we have already shown to be the least efficient star-forming clump in HLS0918. Furthermore, the \citet{nag12-34} models do not account for [CII] dependence on age, or contribution from AGN, and \citet{con02-75} show that N/O (and by extension N/C) may be a poor tracer of metallicity in starbursts and at high redshift, as gas tends to be enriched in primary elements rather than in N. The PDR analysis shows that component VB is diffuse and cooler than the other components, which indicates that VB may be a third galaxy in the group, too distant from Ra--Rb and B to have such vigorous triggered nuclear star-formation.

There is, however, another possible interpretation of component VB. The emission profiles of B and VB are very different, as VB has strong low-$J$ $^{12}$CO lines, as well as relatively bright ionized gas lines, [CI] and [NII]. These pointers, and in particular the high N/CO ratio, indicate that VB may be a high velocity, ionized outflow. Such an outflow could conceivably be driven by the high radiation pressure produced by the intense FIR field that we have already invoked for the H$_2$O emission. Molecular outflows are particularly enticing, as they represent the quenching mechanism in action, expelling fuel from star-forming nuclei. Tentative evidence for a high redshift massive outflows powered by quasar activity has recently been presented for a system at $z=6.4$ \citep{mai12-66}. The situation is no clearer for the $z=4.7$ merger BR1202-0725, which has a quasar host galaxy as a close companion, where two recent analyses of the ALMA [CII] data differ significantly in their conclusions concerning the existence of an outflow \citep[e.g.][]{wag12-30,car13-press}. In HLS0918, the derived conditions are not so extreme as to rule out VB simply being a third galaxy rather than an outflow. Before either scenario can be adopted, further observations are required, such as maps of e.g. OH$^+$, CH$^+$, H$_2$O$^+$. Such species are excellent tracers of molecular outflows, and have been reported for several local ULIRGs, e.g. Arp220, \citep{ran11-94} and Mrk231 \citep{van10-42}.


\section{Conclusions}
\label{sec:conc}

We present new maps of the [CII] (from SMA) and $^{12}$CO(1--0) (VLA) emission for HLS0918, a $z=5.2430$ lensed SMG behind Abell 773. These observations are combined with previously reported line profiles including multiple $^{12}$CO rotational transitions, [CI], water and [NII] \citep{com12-4}. We briefly describe a new lens model, including the contribution from both A773 ($\sim$10\%) and a foreground galaxy at $z=0.62$, and derive a total magnification $\mu_{\rm total}=8.9\pm1.9$. A detailed description of the lens model will be presented in Boone et al. (in prep.). The total FIR luminosity is $L_{\rm FIR,demag} = (1.8\pm0.4)\times 10^{13}$ L$_\sun$, and the updated FIR SED argues for a steep dust emissivity index ($\beta\sim2$). Although HLS0918 is a HyLIRG, the FIR continuum shape resembles that of a local LIRG.

We discover four spectral components to HLS0918 which correspond cleanly to discrete spatial structures identified in the maps. The four components originate from a source-place region separated by $\la4$ kpc, with the three reddest crossing the caustic (double images).

The reddest spectral component (Ra; $V\sim+120\pm30$ km s$^{-1}$) has a de-magnified $L_{\rm FIR,Ra,demag} = (1.1\pm0.2)\times10^{13}$ L$_\sun$, dominating the continuum map, and exhibits a very high star formation efficiency of 360$\pm$90 L$_\sun$ M$_\sun^{-1}$. Both Ra and the second most redshifted component (Rb; $V\sim-130\pm20$ km s$^{-1}$; $L_{\rm FIR,Rb,demag} = (1.9\pm0.4)\times 10^{12}$ L$_\sun$) show strong water emission, possibly excited by the powerful FIR radiation field caused by the intense star formation of Ra, and likely form the nucleus of a massive galaxy.

A third component (B; $-470\pm10$ km s$^{-1}$) is intrinsically twice as luminous as Rb ($L_{\rm FIR,B,demag} = (4.0\pm0.8)\times10^{12}$ L$_\sun$), appearing as an eastern bridge in the continuum map. Water emission is undetected in B, but otherwise the component exhibits a stable luminosity ratio with Rb (Rb/B $\sim1.2\pm0.1$). B is most likely a star-forming region in a second, merging companion to the Ra--Rb galaxy, with a nuclear separation of only $\sim$1 kpc. Components Ra, Rb and B all show evidence of an extended $^{12}$CO(1--0) and [CII] reservoir around the densest star-forming PDR. The elevated [CII] from the diffuse gas explains the observed offset from local ULIRGs: more evolved systems which presumably lack such an extended gas reservoir. On the other hand, the gas properties and morphology of HLS0918 exhibits strong evidence that the Ra--Rb/B system is a late stage merger, consistent with the majority of local ULIRGs.

The faint, bluest component (VB; $V\sim-720\pm40$ km s$^{-1}$; $L_{\rm FIR,VB,demag} = (4.5\pm0.9)\times 10^{11}$ L$_\sun$) originates from a spatially distinct region, which appears to be lower density, cooler and less vigorously forming stars than the other components. The component may also lack an extended molecular gas component, perhaps suggesting a less active companion galaxy which has recently undergone an interaction. However, VB exhibits strikingly bright ionized gas lines ([NII] and [CI]) and a lack of emission in high-$J$ $^{12}$CO transitions. The high N/CO ratio, in particular, could instead indicate that the component originates from a ionized, molecular outflow. Observation of additional species known to trace molecular outflows (e.g. OH$^+$, CH$^+$, H$_2$O$^+$) are required to confirm this scenario.

This comprehensive view of the gas properties and morphology of a system at $z=5.2$ shows the power of gravitational lensing in revealing star formation and galaxy evolution in the early Universe. This paper also offers a preview of the type of analysis that will become possible for a large number of high-redshift galaxies once ALMA progresses to full science operations.

\acknowledgments

TDR is supported by a European Space Agency (ESA) Research Fellowship at the European Space Astronomy Centre (ESAC), in Madrid, Spain. IRS acknowledges support from STFC, a Leverhulme Fellowship, the ERC Advanced Investigator programme DUSTYGAL 321334 and a Royal Society/Wolfson Merit Award.

The authors acknowledge fruitful discussions within the international team lead by D. Schaerer on ``Exploiting the Multi-Wavelength Lensing Survey'' at ISSI (International Space Science Institute) in Bern 2010--2013.

The Submillimeter Array is a joint project between the Smithsonian Astrophysical Observatory and the Academia Sinica Institute of Astronomy and Astrophysics and is funded by the Smithsonian Institution and the Academia Sinica. This work also includes observations carried out with the Karl G. Jansky Very Large Array (VLA): The National Radio Astronomy Observatory is a facility of the National Science Foundation operated under cooperative agreement by Associated Universities, Inc. Additionally, based on observations carried out with the IRAM Plateau de Bure Interferometer and the IRAM 30m Telescope. IRAM is supported by INSU/CNRS (France), MPG (Germany) and IGN (Spain).

This work follows on from observations made with the {\it Herschel} Space Observatory, a European Space Agency Cornerstone Mission with significant participation by NASA. Support for this work was provided by NASA through an award issued by JPL/Caltech. We would also like to thank the HSC and NHSC consortia for support with data reduction.

\end{document}